\documentclass[sigplan,screen]{acmart}

\copyrightyear{2025}
\acmYear{2025}
\setcopyright{acmlicensed}
\acmConference[ASPLOS '25] {Proceedings of the 30th ACM International Conference on Architectural Support for Programming Languages and Operating Systems, Volume 2}{March 30--April 3, 2025}{Rotterdam, Netherlands.}
\acmBooktitle{Proceedings of the 30th ACM International Conference on Architectural Support for Programming Languages and Operating Systems, Volume 2 (ASPLOS '25), March 30--April 3, 2025, Rotterdam, Netherlands}
\acmISBN{979-8-4007-1079-7/25/03}
\acmDOI{10.1145/3676641.3716006}

\usepackage[]{hyperref}
\usepackage{graphicx}
\usepackage{subfigure}
\usepackage{algorithm}
\usepackage{algorithmic}
\usepackage{balance}

\usepackage{subcaption}

\newcommand{\revise}[1]{{\color{black} #1}}

\begin{document}

\title{MoC-System: Efficient Fault Tolerance for Sparse Mixture-of-Experts Model Training}

\author{Weilin Cai}
\affiliation{
  \institution{The Hong Kong University of Science and Technology (Guangzhou)}
  \city{Guangzhou}
  \country{China}
}
\email{wcai738@connect.hkust-gz.edu.cn}

\author{Le Qin}
\affiliation{
  \institution{The Hong Kong University of Science and Technology (Guangzhou)}
  \city{Guangzhou}
  \country{China}
}
\email{lqin674@connect.hkust-gz.edu.cn}

\author{Jiayi Huang}
\authornote{Corresponding author.}
\affiliation{
  \institution{The Hong Kong University of Science and Technology (Guangzhou)}
  \city{Guangzhou}
  \country{China}
}
\email{hjy@hkust-gz.edu.cn}

\begin{abstract}
As large language models continue to scale up, distributed training systems have expanded beyond 10k nodes, intensifying the importance of fault tolerance.
Checkpoint has emerged as the predominant fault tolerance strategy, with extensive studies dedicated to optimizing its efficiency.
However, the advent of the sparse Mixture-of-Experts (MoE) model presents new challenges due to the substantial increase in model size, despite comparable computational demands to dense models.

In this work, we propose the Mixture-of-Checkpoint System (MoC-System) to orchestrate the vast array of checkpoint shards produced in distributed training systems. 
MoC-System features a novel Partial Experts Checkpointing (PEC) mechanism, an algorithm-system co-design that strategically saves a selected subset of experts, effectively reducing the MoE checkpoint size to levels comparable with dense models.
Incorporating hybrid parallel strategies, MoC-System involves fully sharded checkpointing strategies to evenly distribute the workload across distributed ranks.
Furthermore, MoC-System introduces a two-level checkpointing management method that asynchronously handles in-memory snapshots and persistence processes.

We build MoC-System upon the Megatron-DeepSpeed framework, achieving up to a 98.9\% reduction in overhead for each checkpointing process compared to the original method, during MoE model training with ZeRO-2 data parallelism and expert parallelism.
Additionally, extensive empirical analyses substantiate that our methods enhance efficiency while maintaining comparable model accuracy, even achieving an average accuracy increase of 1.08\% on downstream tasks.
\end{abstract}

\ccsdesc[500]{Computer systems organization~Reliability}

\keywords{Fault Tolerance, Checkpoint, Mixture of Experts, Large Language Models, Training}

\maketitle 

\section{Introduction}

Transformer-based large language models (LLMs), which scale to billions or even trillions of parameters, have emerged as the most trending topic in AI research due to their impressive capabilities \cite{achiam2023gpt,chowdhery2023palm,wei2022chain,ouyang2022training,wei2022emergent,brown2020language,attention}. 
Recently, the sparsely-gated mixture-of-experts (MoE) has become the preferred methodology to increase parameter counts and enhance the model quality of LLMs without a proportional increase in computational requirements \cite{ShazeerMMDLHD17,riquelme2021scaling,jiang2024mixtral,cai2024survey}.
To facilitate the training of MoE models and their deployment across expansive computing clusters, distributed training systems have been refined to incorporate expert parallelism (EP) \cite{lepikhin2020gshard,fedus2022switch,hwang2023tutel,singh2023hybrid,he2021fastmoe,cai2024shortcut} alongside established frameworks of data parallelism (DP) \cite{rajbhandari2020zero, ren2021zero} and model parallelism \cite{shoeybi2019megatron, narayanan2021efficient, smith2022using, huang2019gpipe, narayanan2019pipedream, li2021sequence, korthikanti2023reducing}.
With the escalation in the number of deployed computing devices and the incidence of faults \cite{maeng2021understanding,wu2023transom,jiang2024megascale,gupta2017failures,di2014lessons}, ensuring fault tolerance has become a critical component of AI system infrastructure. 

Although numerous studies have effectively addressed fault tolerance for dense (non-MoE) models through periodical checkpoints \cite{wang2023gemini,jiang2024megascale,nicolae2020deepfreeze,wu2023transom,maeng2021understanding,mohan2021checkfreq}, the distinctive characteristics of MoE models necessitate specialized strategies to assure their reliable and efficient fault-tolerant training.
As MoE models scale to unprecedented sizes, the primary challenge in fault tolerance is the substantial increase in checkpoint size, which poses a storage burden that distributed filesystems struggle to handle efficiently \cite{rajbhandari2022deepspeed,shen2022se,du2022glam}. 
Even with prevailing sharded and asynchronous checkpointing strategies \cite{nicolae2020deepfreeze,mohan2021checkfreq,wang2024fastpersist,maurya2024datastates,chen2023cost}, the enlarged checkpointing duration cannot be fully overlapped with the training process, yielding additional costs to the total training time.

Pioneering efficient fault tolerance for MoE model training, we introduce the Mixture-of-Checkpoint System (MoC-System).
The name ``Mixture-of-Checkpoint'' reflects the system's design to orchestrate the vast array of checkpoint shards produced during distributed training.
MoC-System features an innovative Partial Experts Checkpointing (PEC) mechanism, an algorithm-system co-design that reduces the checkpoint size for MoE models by selectively saving only a subset of experts. 
Specifically, PEC selectively saves $K_{pec}$ experts per MoE layer during each checkpointing, while fully saving the non-expert parameters of the model.

It is inspired by observations in MoE model fine-tuning, where updating only the non-expert parameters can achieve the same accuracy as updating all parameters while updating only the expert parameters leads to a drastic reduction in model accuracy \cite{zoph2022st}.
This is believed to be due to the sparsity of the MoE structure, which makes it less sensitive to a limited number of updates, as supported by observations that MoE models generally require larger volumes of pre-training data \cite{artetxe2021efficient, fedus2022switch, xue2022go}.
Building on existing algorithm-system co-design efforts that leverage LLMs' features to optimize computation \cite{dao2022flashattention, liu2024efficient}, communication \cite{rajbhandari2022deepspeed, he2022fastermoe, wang2024towards}, and memory \cite{yi2023edgemoe, hwang2024pre}, we innovatively apply the co-design method to fault tolerance.

Compared to saving the states of all model parameters, using PEC results in a loss of updates to the expert parameters during checkpointing, potentially compromising model accuracy upon recovery.
To quantitatively evaluate the impact of PEC on accuracy, we propose the Proportion of Lost Tokens (PLT) metric, which measures the update loss caused by PEC, as parameter updates are posed by input tokens.
Our empirical results reveal an inverse relationship between model accuracy and PLT, yet we find that the model accuracy maintains akin to the non-fault case when PLT does not exceed 3.75\%.

Building on the efficacy of PEC, MoC-System further introduces the fully sharded checkpointing strategies to distribute workload evenly across distributed ranks, while existing sharding efforts lack specific optimizations for scenarios involving expert parallelism.
Furthermore, MoC-System involves a two-level checkpointing management method that asynchronously controls in-memory snapshot and storage persist processes, with adaptive configuration of hyperparameters for various scenarios.
The refinement of PEC into snapshot-PEC and persist-PEC not only leverages the higher bandwidth of memory and the reliability of distributed storage but also reduces PLT to maintain model accuracy.

We implement the MoC-System and conduct experiments upon the Megatron-DeepSpeed \cite{Megatron_DeepSpeed,smith2022using,rajbhandari2022deepspeed}, which is an acclaimed open-source framework supporting the predominant MoE training strategy of ZeRO-2 DP \cite{rajbhandari2020zero} and EP.
Our experimental results from training the GPT-350M-16E model demonstrate that the PEC approach achieves a 57.7\% reduction in the total checkpoint size. 
Furthermore, recovery from PEC checkpoints maintains comparable validation loss during pre-training and even achieves an average accuracy increase of 1.08\% on downstream tasks.
Additionally, with all optimizations applied, MoC-System reduces overhead for each checkpointing process by up to 98.9\% compared to the original method, and speeds up each training iteration with checkpointing by up to 5.12$\times$.

In summary, our contributions are:
\begin{itemize}
    \item We introduce the Mixture-of-Checkpoint System (MoC-System) to achieve efficient fault tolerance for MoE model training, which integrates multiple strategies to decompose and manage checkpoint shards.
    \item We propose a novel Partial Experts Checkpointing (PEC) mechanism, reducing the checkpoint size by selectively saving a subset of experts. Furthermore, we propose the Proportion of Lost Tokens (PLT) metric to quantitatively assess the accuracy impact of PEC.
    \item We implement the fully sharded checkpointing strategies to distribute workload evenly across distributed ranks, which are applicable to both PEC and conventional checkpointing scenarios.
    \item We design a two-level checkpointing management method that asynchronously handles snapshot and persist processes, maximizing the benefits of PEC.
    \item We conduct extensive experiments to substantiate the superior performance of our approach in enhancing fault tolerance efficiency without sacrificing model accuracy. Additionally, we extend our experiments to examine the impact of varying checkpointed model states, observing that a limited update loss can even improve the accuracy of downstream tasks. 
\end{itemize}

\section{Background \& Related Work}
\subsection{Sparse Mixture-of-Experts (MoE) Models}
The sparse Mixture-of-Experts (MoE) layer \cite{ShazeerMMDLHD17}, consists of multiple feed-forward networks (FFNs), termed ``experts'', and a trainable gating network for selectively activating a subset of these experts. Formally, with $N$ expert networks $\{E_i\}_1^N$, gating network $G$, and input $x$, the MoE layer's output can be formulated as: 
\begin{equation}
\begin{aligned}
MoE(x)=\sum_{i=1}^{N} G(x)_i E_i(x) 
\label{E1}
\end{aligned}
\end{equation}
The common practices in existing MoE research use the noisy top-k softmax gating network to select the top-ranked experts for the computation, formulated as
\begin{equation}
\begin{aligned}
\label{eq:gx}
G(x) = TopK(Softmax(f(x)+\epsilon))
\end{aligned}
\end{equation}
where $f(\cdot)$ denotes the gating linear transformation and $\epsilon$ is the Gaussian noise. 
Leveraging the sparse activations yielded by $G(x)$, this approach facilitates a substantial augmentation of model parameters without causing a proportional increase in computational cost.
Employing the MoE layer to substitute the selected FFN layer in Transformer-based LLMs engenders a significant rise in checkpoint data volume due to the multiplicity of FFN experts, thereby presenting challenges to efficient checkpointing for fault tolerance.

\subsection{Distributed Training of MoE Models}
\label{sec:dist_training}

\begin{figure}
\vspace{-0.10in}
\begin{center}
\centerline{
\begin{minipage}[b]{1\linewidth}
    \subfigure[Model Parameters]{\includegraphics[width=0.496\linewidth]{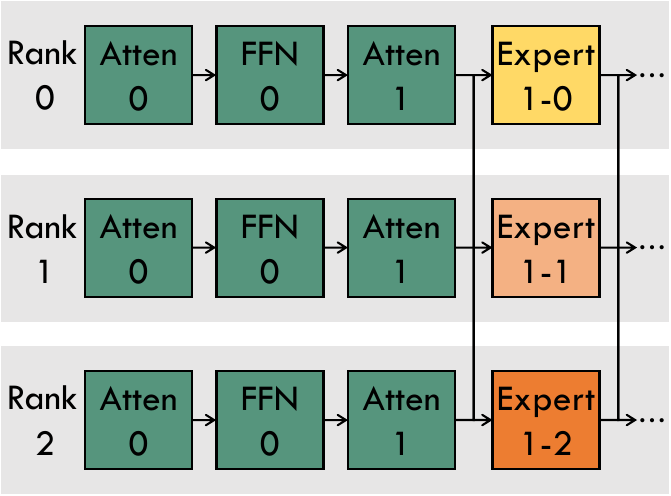}\label{fig:DP_EP_overview_a}}
    \subfigure[Optimizer States]{\includegraphics[width=0.496\linewidth]{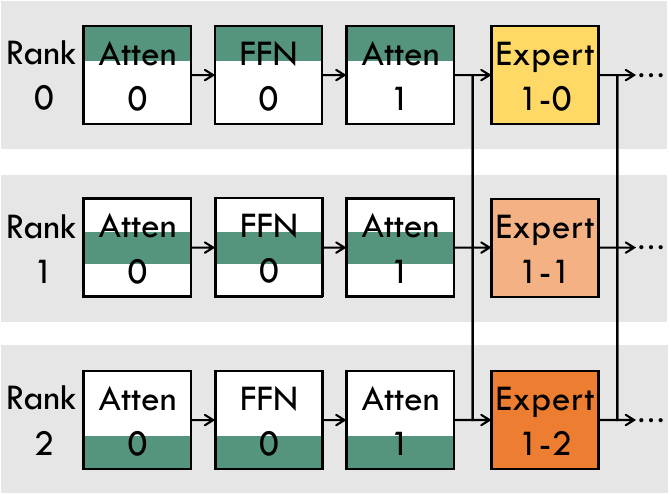}\label{fig:DP_EP_overview_b}}
\end{minipage}
}
\vspace{-0.15in}
\caption{An illustration of the model states, including model parameters (a) and optimizer states (b), across three ranks in distributed training. The training utilizes the hybrid parallel strategy of ZeRO-2 DP + EP, configured with the parallel degree of DP = 3 and EP = 3. The non-expert parts are depicted in green, while the expert parts are depicted in yellow, with varying shades denoting different experts within the same MoE layer. The combination of white and green in the non-expert modules in (b) illustrates the partitioning of states across ranks through ZeRO-2 DP. ``Atten0'' and ``FFN0'' represent Attention and FFN sublayers in the $0th$ transformer layer, while ``Atten1'' and the MoE layer, including ``Expert(1-0, 1-1, 1-2)'', are in the $1th$ transformer layer.}
\label{fig:DP_EP_overview}   
\end{center}
\vspace{-0.2in}
\end{figure}


The adoption of MoE in LLMs introduces new challenges to existing training and inference systems, due to its inherently sparse and dynamic computational workload.
GShard \cite{lepikhin2020gshard} pioneers the parallel strategy of Expert Parallelism (EP) by facilitating parallel gating and expert computation. Specifically, EP assigns distinct experts to each distributed computing device such as GPU and TPU, and passes input tokens to the corresponding experts via All-to-All communication. 
Following this, EP has ascended as a pivotal strategy, enabling the efficient scaling of MoE model training \cite{fedus2022switch, rajbhandari2022deepspeed, hwang2023tutel, singh2023hybrid}.

As depicted in Figure~\ref{fig:DP_EP_overview_a}, EP can be viewed as an augmentation of Data Parallelism (DP) \cite{rajbhandari2020zero, ren2021zero, rajbhandari2021zero}, where each expert within an MoE layer is allocated to a distinct DP rank (e.g., ``Expert1-0'' on ``Rank0'' and ``Expert1-1'' on ``Rank1''), while all non-expert layers (e.g., ``Atten0'', ``FFN0'', and ``Atten1'') are replicated across DP ranks. 
Moreover, the synergy of EP with other parallel strategies, such as Tensor Parallelism (TP) \cite{shoeybi2019megatron, smith2022using, narayanan2021efficient}, Pipeline Parallelism (PP) \cite{huang2019gpipe, narayanan2019pipedream, qi2023zero}, has been explored to enhance the scalability and efficiency of MoE model training in expansive distributed settings \cite{fedus2022switch,singh2023hybrid,hwang2023tutel,zhai2023smartmoe,he2022fastermoe,wei2024skywork}. 
From the checkpoint perspective, a notable distinction between EP and other parallelism is EP’s flexibility in distributing diverse parameters across DP ranks. In contrast, TP and PP maintain parameters replicated across all DP ranks, limiting their adaptability within each DP rank.

In this work, we primarily focus on distributed training with the hybrid parallel strategy of ZeRO-2 DP + EP (notably, ZeRO-1 is analogous to ZeRO-2 from the view of checkpointing \cite{rajbhandari2020zero}), which has emerged as the predominant approach for training MoE models \cite{rajbhandari2022deepspeed,llama-moe-2023,wei2024skywork}. 
This approach is highlighted for its accessibility and efficiency, supported by Megatron-DeepSpeed \cite{Megatron_DeepSpeed,smith2022using}, an acclaimed open-source distributed training framework. 
Moreover, extensive practical experience with large-scale distributed systems has demonstrated its superior performance, minimizing communication overhead while remaining memory-efficient \cite{rajbhandari2020zero,rajbhandari2022deepspeed,cai2024survey,SimAI}.
Additionally, our proposed checkpointing techniques can be seamlessly extended to other hybrid parallel strategies, encompassing TP and PP, as they can be viewed as the modularity of each DP rank.


\subsection{Fault-tolerant Checkpointing for Distributed Training System}
\label{sec:overhead}

\begin{figure}
    \centering
    \includegraphics[width=1\linewidth]{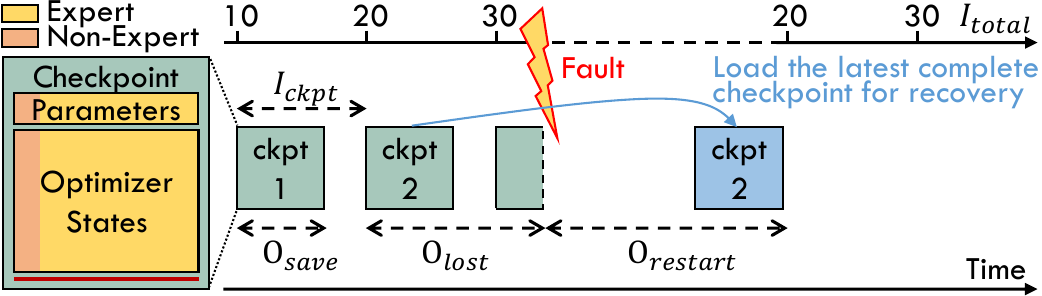}
    \vspace{-0.3in}
    \caption{An illustration of fault tolerance in model training through checkpoint mechanism. The checkpointing interval $I_{ckpt}$ is set to 10 iterations. A fault arises following the 30th iteration, before the completion of the third checkpoint. Therefore, the most recent completed checkpoint (ckpt2) is loaded to recover the training progress. The composition of a checkpoint is depicted on the left, with the size of each component reflecting its data volume, using the GPT-350M-16E model as an example.
    }
    \vspace{-0.15in}
    \label{fig:ft_process}
\end{figure}

Checkpoint serves as a critical mechanism for augmenting fault tolerance in distributed training systems by facilitating the periodic preservation and recovery of model states during training \cite{mohan2021checkfreq,jiang2024megascale,wu2023transom,le2023bloom,koo1987checkpointing}. 
As illustrated in Figure~\ref{fig:ft_process}, the saved model states at each checkpoint comprise learnable model parameters for the expert part (12\% of the total volume) and that for the non-expert part (2\%), optimizer states for the expert part (74\%) and that for the non-expert part (12\%), along with other crucial states (less than 1\%), such as epoch/iteration numbers and Random Number Generator states.
The checkpoint ensures that training progress is not lost in unexpected faults and can be recovered after a restart. 

However, the checkpointing process incurs significant data transfer and storage overhead, alongside additional overhead in the event of a fault.
The total overhead introduced by fault tolerance with checkpoint during the entire model training, denoted as $O_{ckpt}$, can be quantified by aggregating the overhead of a checkpointing process (saving model states) $O_{save}$ during normal training, the overhead of system/task restart $O_{restart}$ and lost training progress $O_{lost}$ when a fault occurs, as illustrated in Figure~\ref{fig:ft_process}. It is formulated as:
\begin{equation}
\begin{aligned}
\label{eq:overhead_1}
O_{ckpt} = O_{save}\frac{I_{total}}{I_{ckpt}} + \sum\nolimits_{i=1}^{N_{fault}} (O_{restart}^i + O_{lost}^i)
\end{aligned}
\end{equation}
where $I_{total}$ represents the total number of iterations in training and $I_{ckpt}$ denotes the iteration interval of checkpointing. 
Each fault occurrence, totaling $N_{fault}$, contributes to the overhead, with $O_{lost}$ being contingent on $I_{ckpt}$ and averaging $\frac{I_{ckpt}}{2}$, and $O_{restart}$ remaining relatively constant. Therefore, the overhead of fault tolerance be represented roughly as the following formulation:
\begin{equation}
\begin{aligned}
\label{eq:overhead_2}
O_{ckpt} \approx O_{save}\frac{I_{total}}{I_{ckpt}} + \sum\nolimits_{i=1}^{N_{fault}} (O_{restart} + \frac{I_{ckpt}}{2})
\end{aligned}
\end{equation}
It is obvious that $I_{ckpt}$, $O_{save}$, and $O_{restart}$ are the key factors determining the total overhead $O_{ckpt}$.
In pursuit of optimizing fault tolerance efficiency, existing research has explored various methods to diminish the above factors.

\begin{figure}
    \centering
    \includegraphics[width=1\linewidth]{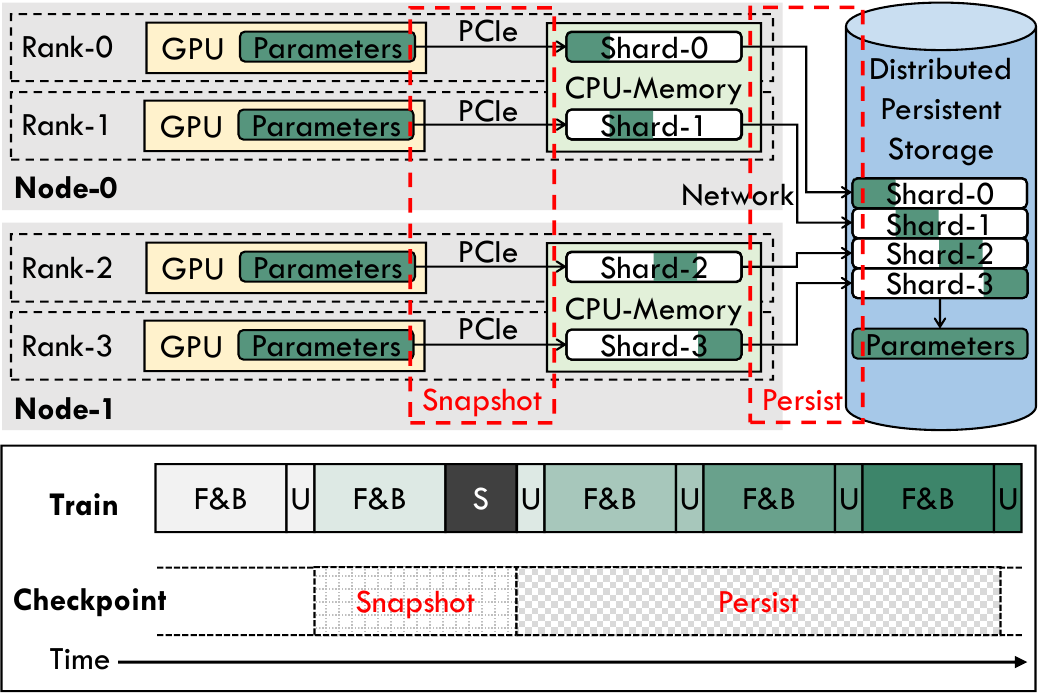}
    \caption{The top half part illustrates the two-phase checkpointing workflow (GPU-to-CPU snapshot + CPU-to-Storage persist) during a distributed training. The training employs 4-degree DP across two nodes, each equipped with two GPUs. Data-parallel sharding is utilized to minimize the volume of data saved per DP rank. The bottom half part presents a timeline for asynchronous checkpointing, where ``F\&B'' denotes the forward and backward passes of an iteration, ``U'' denotes a weight update, ``S'' denotes a checkpoint stall.}
    \label{fig:two_phase_overview}
\end{figure}

\subsubsection{Two-phase Asynchronous Checkpointing.}
\label{sec:background_async}
As demonstrated in Figure~\ref{fig:two_phase_overview}, checkpointing model states from GPU memory to distributed persistent storage involves two phases: transferring from GPU memory to CPU memory (GPU-to-CPU snapshot) and from CPU memory to distributed persistent storage (CPU-to-Storage persist). 
The CPU-to-Storage persist phase involves serializing the model states and writing them to a distributed filesystem via the network, while the GPU-to-CPU snapshot phase copies tensors through PCIe. 
Given that both phases can significantly hinder training progress if executed in a blocking manner, asynchronously processing and overlapping them with ongoing training has emerged as a critical method to enhance checkpointing efficiency \cite{mohan2021checkfreq,wang2023reliable,jiang2024megascale,wan2024bytecheckpoint,wang2024fastpersist,maurya2024datastates,chen2023cost}.

The timeline illustrated in Figure~\ref{fig:two_phase_overview} indicates that the asynchronous GPU-to-CPU snapshot can proceed concurrently with the forward and backward passes (denoted as ``F\&B'') of the subsequent iteration, although it must finish before the weight update phase. 
If the snapshot duration exceeds the ``F\&B'' period, it will trigger a checkpoint stall (``S''), thereby stopping the training process until the snapshot completion. 
Unlike the GPU-to-CPU snapshot, the CPU-to-Storage persist phase is not subjected to this limitation, as the snapshots residing in CPU memory can remain unaffected by the ongoing training process.

However, existing asynchronous checkpointing systems face the new challenges posed by MoE models:
(1) MoE models extend the checkpointing duration without a corresponding increase in ``F\&B'' time, resulting in incomplete overlap of the GPU-to-CPU snapshot and potential checkpoint stalls; (2) prolonged CPU-to-Storage persist leads to an enlarged $I_{ckpt}$. 
In contrast, our methodologies effectively manage the volumes of data transferred during both the snapshot and persist phases, thereby addressing these issues.

\subsubsection{Data-Parallel Sharding.}
\label{sec:DPSharding}
Considering that data volume determines the duration of communication and storage, eliminating redundancies and reducing checkpoint size through data-parallel sharding \cite{nicolae2020deepfreeze,wang2023reliable,wan2024bytecheckpoint,lian2024universal} is an effective optimization. 
Since the original DP replicates model states across all the DP ranks, each rank can store a distinct sharding of the states, collectively forming the complete model states through the aggregation of all ranks' shards, as depicted in Figure~\ref{fig:two_phase_overview}.

With the evolution of DP techniques, model states may already be uniformly partitioned across each DP rank. 
For instance, ZeRO-1 and ZeRO-2 DP \cite{rajbhandari2020zero} partition the optimizer states, whereas ZeRO-3 DP and Fully Sharded Data Parallel (FSDP) \cite{zhao2023pytorch} partition both the model parameters and optimizer states. 
As discussed in Section~\ref{sec:dist_training}, our focus is on the ZeRO-2 DP + EP scenario, where model parameters are replicated across each DP rank, as illustrated in Figure~\ref{fig:DP_EP_overview}.

However, existing distributed training frameworks lack an efficient data-parallel sharding strategy for MoE model training. 
For instance, the Megatron-DeepSpeed framework \cite{Megatron_DeepSpeed} confines the checkpointing of expert model states to the first EP group, as illustrated by Figure~\ref{fig:fully_sharded_checkpointing_a}, neglecting the potential of distributed sharding across all EP groups. 
In contrast, we implement fully sharded checkpointing for MoE model training and further introduce an adaptive sharding strategy with our PEC mechanism, outperforming the commonly used equal sharding strategy.

\subsubsection{In-memory Checkpointing.} 
Due to the superior bandwidth of GPU-to-CPU copy and compute network data transfers compared to CPU-to-Storage persist, several studies minimize $O_{save}$ by opting to store model states in the CPU memory of other nodes instead of persistent storage \cite{wang2023gemini,wang2023reliable}. 
This approach significantly reduces the duration of checkpointing, thereby allowing for lower $I_{ckpt}$ and $O_{ckpt}$.

However, the in-memory checkpointing solution encounters reliability problems within real-world large-scale GPU clusters \cite{wan2024bytecheckpoint}. 
In such environments, multiple nodes within the same backup group may experience simultaneous failures, resulting in severe data loss. 
In contrast, we introduce a two-level checkpoint management strategy that benefits from the efficiency of in-memory checkpointing while ensuring fault tolerance across a wide range of scenarios.

\begin{figure}
    \centering
    \includegraphics[width=1\linewidth]{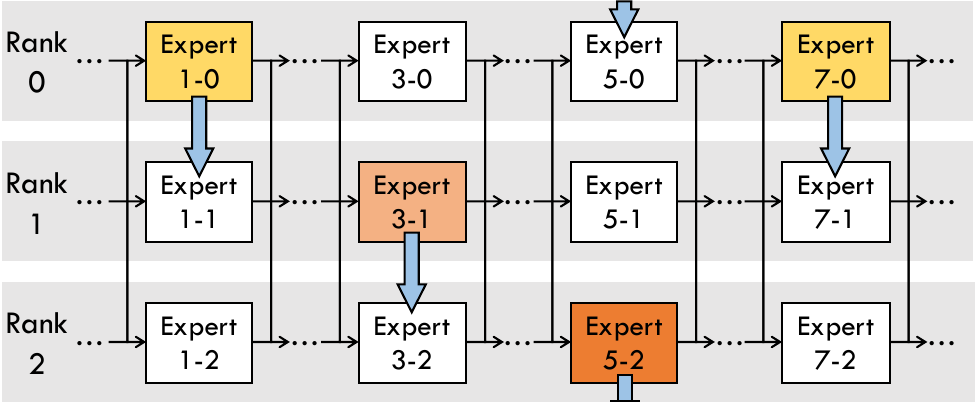}
    \vspace{-0.3in}
    \caption{An illustration of our proposed partial experts checkpointing (PEC) with sequential selection. At the current checkpointing, ``Expert(1-0, 3-1, 5-2, 7-0)'' are saved, while those not saved are marked in white. Blue arrows indicate the iterative pattern of the sequential selection, which will save ``Expert(1-1, 3-2, 5-0, 7-1)'' at the next checkpointing.}
    \vspace{-0.1in}
    \label{fig:pec}
\end{figure}

\subsubsection{Partial Checkpointing and Recovery.} 
Previous research \cite{qiao2019fault} has demonstrated that the iterative-convergent nature of machine learning (ML) training is capable of compensating for the inconsistencies introduced by partial checkpointing and recovery to some extent on Parameter Server (PS) distributed training scenarios. 
Given that the model parameters are distributed across multiple PS nodes, a partial failure of these nodes is likely to result in only a partial loss of the updates to the model parameters. 
Compared to the process of saving and reloading the entire model states, partial checkpointing and recovery strategies can significantly decrease the data volume required for checkpoints. 
This approach has subsequently proven to be effective in the training of the Deep Learning Recommendation Model (DLRM) \cite{naumov2019deeplearningrecommendationmodel,maeng2021understanding,eisenman2022check}, which only accesses and updates a small segment of the model in each iteration. 

However, large-scale distributed training systems and their distributed parallel strategies employed by Transformer-based LLMs differ significantly from PS and DLRM scenarios. 
Consequently, no work has yet explored the integration of partial methods into the fault tolerance of LLMs. 
Our work pioneers in identifying the synergy between the inherent sparsity of MoE LLMs and partial strategies, leading to the development of a more efficient fault tolerance method without harming the final model quality.

\section{Partial Experts Checkpointing}
\label{sec:pec}

In light of the substantial increase in checkpoint size predominantly attributed to the multiplicity of FFN experts within the MoE model, we introduce the concept of Partial Experts Checkpointing (PEC) to significantly downsize the checkpoint.
In the PEC approach, a subset of experts---specifically, $K_{pec}$ of the $N$ experts per MoE layer---is saved, while the non-expert parts are preserved in their entirety. This strategy results in a checkpoint size comparable to that of a dense model when $K_{pec}$ is set to 1, as illustrated in Figure~\ref{fig:pec}. As a device-agnostic checkpointing mechanism, PEC is generally applicable across various MoE model training scenarios.

\begin{figure}
\vspace{-0.10in}
\begin{center}
\centerline{
\begin{minipage}[b]{1\linewidth}
    \subfigure[PLT (\%)]{\includegraphics[width=0.496\linewidth]{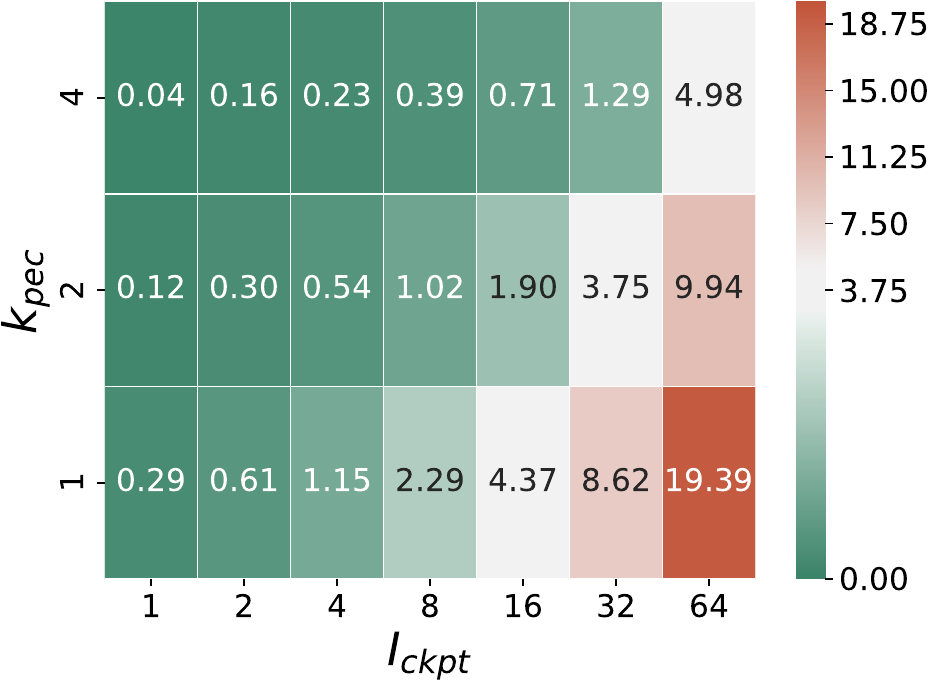}}
    \subfigure[Validation loss]{\includegraphics[width=0.496\linewidth]{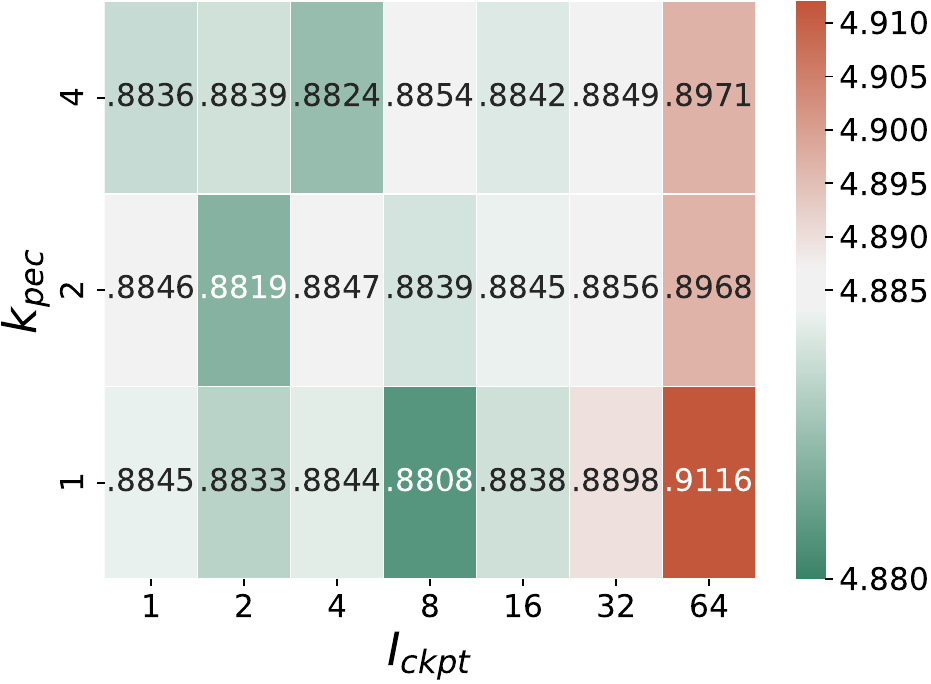}}
\end{minipage}
}
\vspace{-0.15in}
\caption{Correlation analysis between (a) the Proportion of Lost Tokens (PLT) and (b) the final validation loss.
In (a), the PLT centers on 3.75\% observed in a PEC configuration of $K_{pec}=2$ and $I_{ckpt}=32$, which slightly degrades the model accuracy compared to the non-fault case. The validation losses are presented in (b), where the non-fault case's loss of 4.8851 is taken as the center value to highlight the accuracy deviations under various PEC configurations.}
\label{fig:plt}   
\end{center}
\vspace{-0.2in}
\end{figure}

\subsection{Analysis} 
\subsubsection{Checkpoint Size}
To accurately assess the efficacy of PEC on reducing checkpoint size, we initially define the size of a conventional checkpoint, denoted as $C_{full}$, which saves the states of all model parameters. The formulation is as follows:
\begin{equation}
\begin{aligned}
C_{full} \approx (P_{ne} + P_{e}) \cdot (B_{w} + B_{o}) 
\label{eq:ckpt_size}
\end{aligned}
\end{equation}
where $P_{ne}$ and $P_{e}$ denote the number of parameters in the non-expert and expert parts of the model, respectively. Each parameter contributes fixed bytes of weight ($B_{w}$) and optimizer state ($B_{o}$).

PEC, by contrast, only saves a subset of experts at each checkpoint, leading to a reduced checkpoint size, denoted as $C_{pec}$, which is formulated as:
\begin{equation}
\begin{aligned}
C_{pec} \approx (P_{ne} + \frac{K_{pec}}{N}P_{e}) \cdot (B_{w} + B_{o}) 
\label{eq:ckpt_pec_size}
\end{aligned}
\end{equation}
where $K_{pec}$ denotes the number of experts saved per MoE layer, and $N$ denotes the total number of experts in each MoE layer.
Given that the expert part typically constitutes the majority of the model parameters in existing MoE models, PEC's capability to reduce checkpoint size is considerable.

\subsubsection{Impact on Model Accuracy}
\label{sec:PEC_analysis_accuracy}
It is critical to consider that recovering training from a PEC checkpoint may impact the model accuracy, as it causes a loss of expert updates contributed by the training input tokens.
Specifically, the recovery process can retrieve the latest model states of the non-expert part and $K_{pec}$ experts from the latest checkpoint, while the remaining $N-K_{pec}$ experts can only be recovered to their states saved at the previous checkpointing.

To quantitatively assess the potential impact on accuracy attributed to PEC, we introduce a novel metric, the \textbf{Proportion of Lost Tokens (PLT)}. 
The PLT metric is designed to quantify the average proportion of tokens lost across all the MoE layers throughout the training, formulated as follows:
\begin{equation}
\begin{aligned}
PLT=\frac{1}{N_{moe}} \sum_{i=1}^{N_{moe}} \frac{\sum_{j=1}^{N_{fault}} L_{i,j}(I_{ckpt}, K_{pec}, F)}{T_{i} \cdot TopK_{i}} 
\label{plt}
\end{aligned}
\end{equation}
where $N_{moe}$ denotes the number of MoE layers within the model, and $N_{fault}$ denotes the count of faults encountered during the training.
$L_{i,j}$ refers to the measured number of the $ith$ MoE layer's lost tokens caused by the $jth$ fault, which is influenced by the checkpointing interval $I_{ckpt}$, the number $K_{pec}$ of saved experts, and the function $F$ for partial experts selection (e.g. sequential or load-aware methods).
The product of the number of input tokens $T_{i}$ and $TopK_{i}$ of MoE gating indicates the total number of tokens processed by all experts in the $ith$ MoE layer during the training. 
It is worth noting that the actual count of tokens processed by all experts is typically less than $T_{i} \cdot TopK_{i}$, primarily due to the token dropout imposed by the expert capacity \cite{lepikhin2020gshard}.

To investigate the correlation between PLT and model accuracy, we conduct experiments that train GPT-125M-8E models with varying PEC configurations (different values of $K_{pec}$ and $I_{ckpt}$) on Wikitext dataset \cite{wikitext}. 
Each model's training process is designed to encounter a fault at the midpoint, followed by a recovery from the saved PEC checkpoint.

As evidenced in Figure~\ref{fig:plt}, the final validation loss of the models experiences fluctuations (4.8808-4.8856) yet remains comparable to the non-fault case (4.8851) when PLT is below 3.75\%.
It substantiates the efficacy of PEC in minimizing checkpoint size without harming model accuracy in the case of limited PLT (more comprehensive accuracy evaluations are in Section~\ref{sec:eval_accuracy}). 
Additionally, the results highlight a correlation between smaller $K_{pec}$ and larger $I_{ckpt}$ with increased PLT, which may impact the model accuracy.

\subsection{Partial Experts Selection} 

As PEC only saves a subset of experts at each checkpoint, selecting which experts to save is important. 
Different functions of partial expert selection can lead to variations in PLT and the recovered model states, thereby potentially impacting the final model accuracy.

More importantly, considering that experts within each MoE layer are distributed across various ranks and devices via EP, as depicted in Figure~\ref{fig:pec}, the partial expert selection significantly affects the workload distribution across ranks, thus impacting the time cost of checkpointing.
The most imbalanced workload scenario, for instance, involves checkpointing ``Expert(1-0, 3-0, 5-0, 7-0)'', all located in ``Rank0''. 
In this case, the checkpointing duration is primarily prolonged by ``Rank0'', which bears the heaviest workload.

\textbf{Sequential Selection.}
Given the challenges associated with the selection of partial experts, we propose a sequential selection strategy that sequentially alternates the target experts, incorporating an interleaved schedule across MoE layers and EP ranks.
For instance, as illustrated in Figure~\ref{fig:pec}, at the first checkpointing time, ``Rank0'' saves ``Expert(1-0, 7-0)'', ``Rank1'' saves ``Expert3-1'', and ``Rank2'' saves ``Expert5-2''. 
At the next checkpointing time, ``Rank0'' saves ``Expert5-0'', ``Rank1'' saves ``Expert(1-1, 7-1)'', and ``Rank2'' saves ``Expert3-2''. 
With this strategy, PEC can achieve a balanced checkpointing workload while maintaining an acceptable PLT.

\textbf{Load-aware Selection.} 
We extend our investigation into the function of partial expert selection by incorporating a load-aware approach that prioritizes the checkpointing of $K_{pec}$ experts, characterized by the highest number of unsaved updates.
Based on our empirical results, load-aware selection achieves model accuracy on par with sequential selection but necessitates more complicated control mechanisms and incurs higher costs, making it a less favorable option.

\section{Fully Sharded Checkpointing}
\label{sec:FCS}

As discussed in Section~\ref{sec:DPSharding}, existing work lacks an efficient data-parallel sharding strategy for checkpointing MoE models in distributed training. Figure~\ref{fig:fully_sharded_checkpointing_a} demonstrates that the baseline method provided by the Megatron-DeepSpeed framework \cite{Megatron_DeepSpeed} only utilizes ``Rank0'' to save non-expert states and ``EP-Group-0'' to save expert states.

\subsection{Equal Sharding for Expert Part}
\begin{figure}
\vspace{-0.10in}
\begin{center}
\centerline{
\begin{minipage}[b]{1\linewidth}
    \subfigure[Model Parameters]{\includegraphics[width=0.496\linewidth]{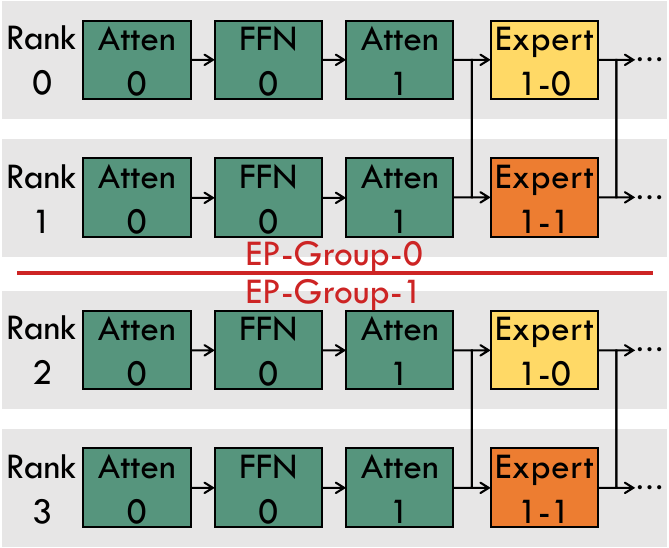}\label{fig:DP4EP2_a}}
    \subfigure[Optimizer States]{\includegraphics[width=0.496\linewidth]{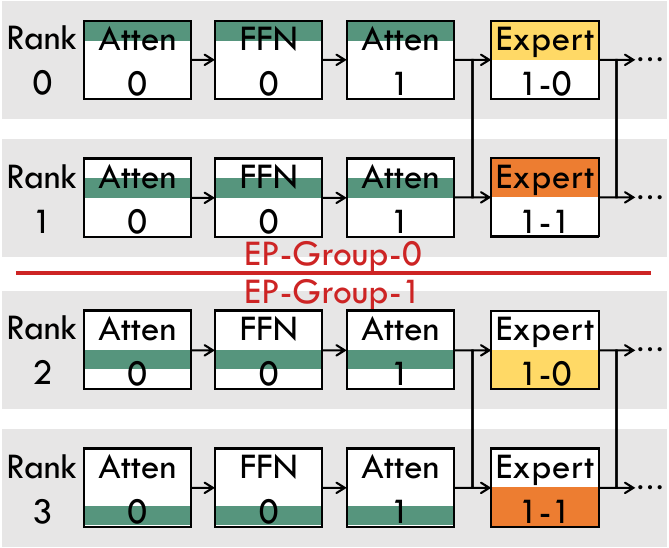}\label{fig:DP4EP2_b}}
\end{minipage}
}
\vspace{-0.15in}
\caption{An illustration of the model states across 4 distributed ranks in training. The training utilizes the hybrid parallel strategy of ZeRO-2 DP + EP, configured with the parallel degree of DP = 4 and EP = 2.}
\label{fig:DP4EP2}   
\end{center}
\vspace{-0.2in}
\end{figure}

In contrast to the hybrid strategy depicted in Figure~\ref{fig:DP_EP_overview}, which employs a single EP group, the prevailing practice \cite{liu2024deepseek,wei2024skywork} for large-scale distributed training of MoE models employs multiple EP groups, as demonstrated in Figure~\ref{fig:DP4EP2}. 

To enhance efficiency by evenly distributing the checkpointing workload across distributed ranks, we implement an equal sharding strategy for the expert part of the MoE model. 
This strategy employs each expert as the smallest unit for distribution across various EP groups, each containing replicas of the same experts. 
As exemplified in Figure~\ref{fig:fully_sharded_checkpointing}, ``Rank0'' in ``EP-Group-0'' is allocated the first half of ``Expert0'', while ``Rank2'' in ``EP-Group-1'' is assigned the second half.

\subsection{Equal Sharding for Non-Expert Part}
\begin{figure}
\vspace{-0.10in}
\begin{center}
\centerline{
\begin{minipage}[b]{1\linewidth}
    \subfigure[Baseline]{\includegraphics[width=0.496\linewidth]{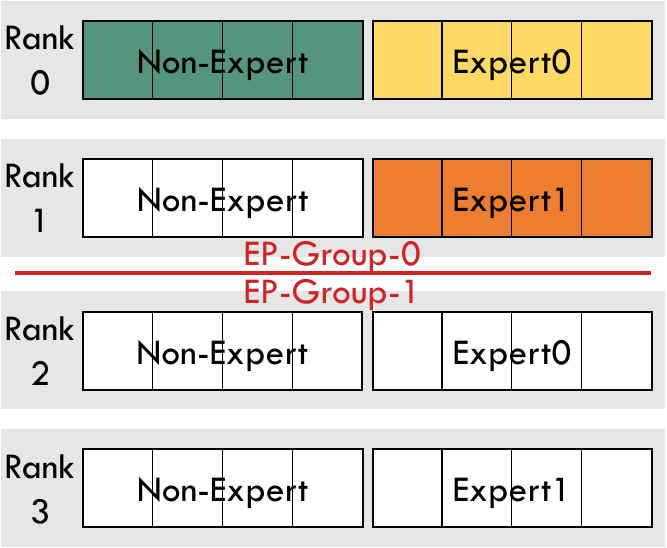}\label{fig:fully_sharded_checkpointing_a}}
    \subfigure[Our Fully Sharded Checkpointing]{\includegraphics[width=0.496\linewidth]{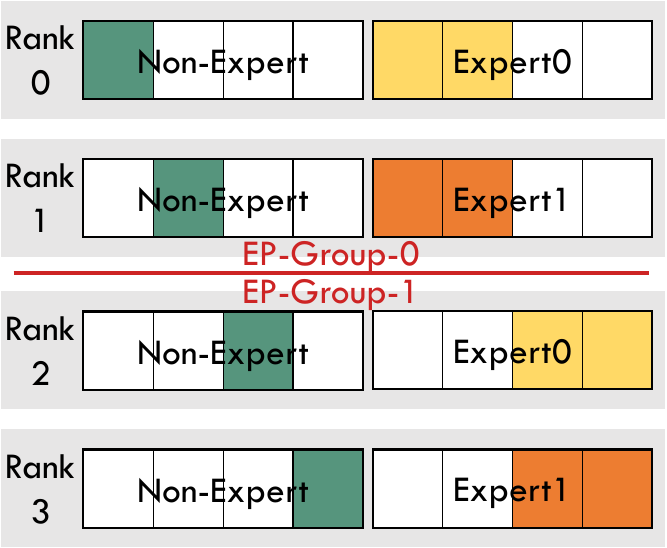}\label{fig:fully_sharded_checkpointing_b}}
\end{minipage}
}
\vspace{-0.15in}
\caption{An illustration of two distinct checkpointing methods employed for training the MoE model, configured with DP = 4 and EP = 2. 
    (a) illustrates the baseline method provided by the Megatron-DeepSpeed framework. (b) presents our proposed fully sharded checkpointing with equal sharding.
    For simplification, the model states are divided into two segments: the non-expert and the expert parts. 
    The horizontal segments within each part represent various layers.
    }
\label{fig:fully_sharded_checkpointing}   
\end{center}
\vspace{-0.2in}
\end{figure}

Given the considerable overhead associated with fine-grained sharding methods \cite{nicolae2020deepfreeze} and the fact that the model parameters of the non-expert part comprise only 2\% of the total checkpoint volume, we implement a coarse-grained sharding approach, utilizing layers (e.g. Attention and FFN) as the minimum partition units. 
Building upon this framework, we introduce an equal sharding strategy, aiming to evenly distribute the workload of checkpointing non-expert layers across all DP ranks, as depicted in Figure~\ref{fig:fully_sharded_checkpointing_b}.
We define the ideal checkpointing workload of each rank, $C_{rank}$, using the following formulation:
\begin{equation}
\begin{aligned}
C_{rank} \approx \frac{(P_{ne} + P_{e}) \cdot B_{o}}{D_{ep}} + \frac{P_{ne} \cdot B_{w}}{D_{dp}} + \frac{P_{e} \cdot B_{w}}{D_{ep}}
\label{eq:ckpt_size_fsc}
\end{aligned}
\end{equation}
where $D_{dp}$ and $D_{ep}$ denote the parallel degree of DP and EP, respectively.
While this method may not achieve exact equality as observed in tensor-level sharding, it markedly diminishes the control cost.
Additionally, the sharding pattern of each rank is established during the initial stage and maintained throughout the training.

\subsection{Adaptive Sharding for Non-Expert Part}
\label{sec:adaptive_sharding}
PEC may lead to an imbalanced checkpointing workload for the expert part if the following conditions are met:
\begin{equation}
\begin{aligned}
(K_{pec} \cdot N_{moe}) \bmod D_{ep} \neq 0 \quad \text{or} \quad \frac{K_{pec} \cdot N_{moe}}{D_{ep}} \bmod \frac{D_{dp}}{D_{ep}} \neq 0.
\label{eq:ckpt_size_fsc}
\end{aligned}
\end{equation}
Using Figure~\ref{fig:pec} as an example, ``Rank0'' is responsible for saving two experts, whereas the other ranks save one expert each, resulting in an imbalanced workload.

To leverage the spare capacity across ranks, we introduce an adaptive sharding strategy, which adaptively allocates non-expert parts based on the selection pattern of PEC. 
Furthermore, it incorporates a greedy algorithm for shard allocation, prioritizing the assignment of larger modules to ranks exhibiting the least accumulated workload. 
Additionally, the initially established sharding pattern can also be consistently applied throughout the training process, without the need for further synchronization or dynamic adjustments at runtime, due to the consistency of the PEC sequential selection.

In our implementation, sharding strategies are exclusively utilized to partition model parameters, tailored to our specific scenario of ZeRO-2 DP + EP, where optimizer states are already partitioned, as depicted in Figure~\ref{fig:DP4EP2}. 
Nevertheless, our methodologies are applicable to the partitioning of both model parameters and optimizer states in scenarios that do not incorporate ZeRO sharding \cite{rajbhandari2020zero}.


\section{Two-Level Checkpointing Management}

\begin{figure*}
    \centering
    \includegraphics[width=1\linewidth]{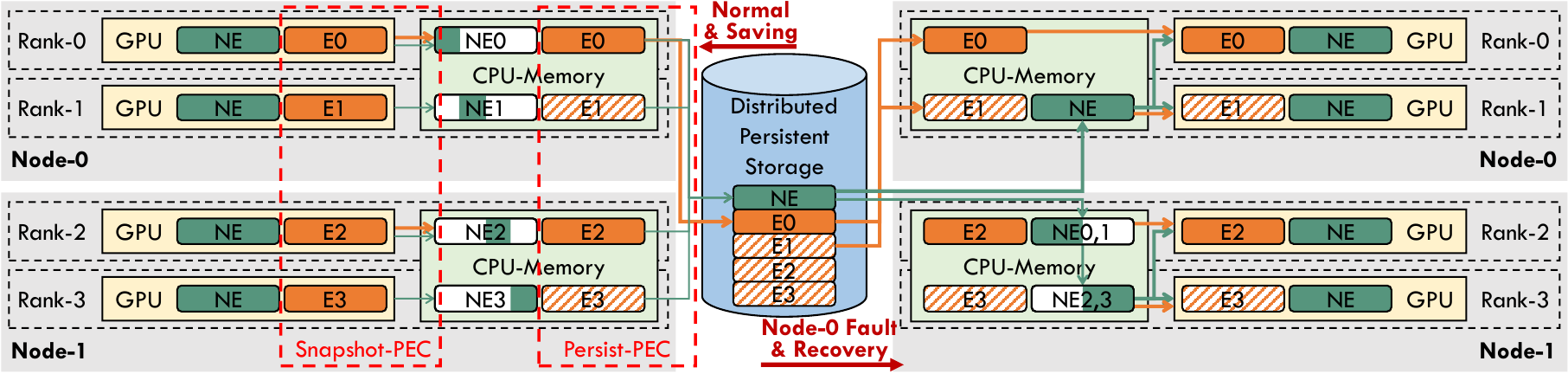}
    \vspace{-0.25in}
    \caption{Illustrations of the saving process during normal training (left half) and the recovery process after a fault occurrence (right half), as implemented in our two-level checkpointing management. Orange ``E(0-3)'' denote distributed expert parts, while green ``NE'' denotes the non-expert part, and ``NE(0-3)'' denote the ``NE'' shards. The data transfer for expert and non-expert parts is represented by arrows in matching colors. The diagonal filled ``E(0-3)'' indicates not involved at the latest checkpoint.}
    \label{fig:system}
    \vspace{-0.05in}
\end{figure*}

To maximize the benefits of hierarchical storage, we propose a two-level checkpointing management into our MoC system, comprising (1) in-memory snapshot and (2) persist, coupled with a suite of optimization techniques. 

\subsection{Two-level PEC Saving and Recovery}
As depicted in Figure~\ref{fig:system}, we implement the saving and recovery processes across CPU memory and storage, which takes advantage of the superior GPU-to-CPU bandwidth and distributed storage's persistence.

\textbf{Saving.} We introduce the snapshot-PEC and persist-PEC processes, designed to alleviate data transmission burdens during their respective levels. 
Furthermore, we refine the hyperparameter $K_{pec}$ in PEC into two variables: $K_{snapshot}$ and $K_{persist}$. 
This distinction allows snapshot-PEC to select $K_{snapshot}$ out of $N$ experts for transfer from GPU to CPU memory. 
Concurrently, persist-PEC is tasked with selecting $K_{persist}$ out of the $K_{snapshot}$ experts for subsequent storage persistence.
As exemplified in Figure~\ref{fig:system}, snapshot-PEC saves only E(0,1) to CPU memory, followed by persist-PEC, which saves only E0 to storage.
To streamline the management of all checkpointed model modules, we utilize key-value pairs for efficient retrieval from both memory and distributed storage. 

\textbf{Recovery.} 
In the event of a fault, while the fault nodes may lose their data, other normal nodes can recover from the in-memory snapshots, thus not only reducing the overhead of loading data from persistent storage but also mitigating the PLT attributable to persist-PEC. 
Take Figure~\ref{fig:system} as an example, the restarted Node-0 needs to load NE and E(0,1) from persistent storage, while Node-1 only needs to load NE(0,1).
Furthermore, Node-1 benefits from recovering E(2,3) directly from memory, which contains more recent states than those available in storage, thereby reducing the PLT.

\begin{figure}
\centering
\includegraphics[width=\linewidth]{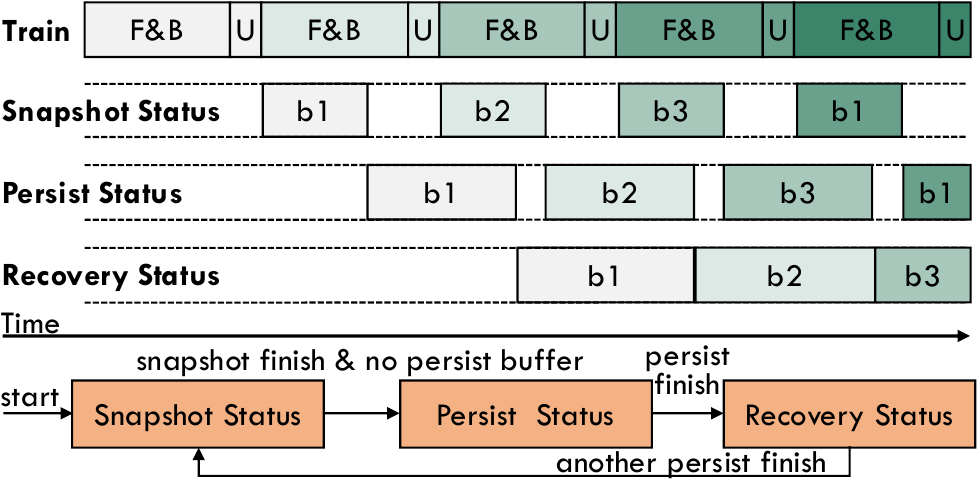}
\vspace{-0.2in}
\caption{The timeline of the asynchronous checkpointing process with triple-buffer. The orange part illustrates the status transition among the three buffers (``b1'', ``b2'', ``b3''). The time span of each buffer in a snapshot or persist status can reflect the time cost of the snapshot or persist process.}
\vspace{-0.0in}
\label{fig:timeline}
\end{figure}

\subsection{Asynchronous Checkpointing \& Triple Buffering}
To minimize the overhead of states saving $O_{save}$, we implement an asynchronous checkpointing mechanism that allows checkpointing to overlap with the normal training processes.
Specifically, we develop an agent at each node to facilitate the two-level checkpointing management through a triple-buffer mechanism.
As illustrated in Figure~\ref{fig:timeline}, the triple buffering comprises snapshot, persist, and recovery buffers, each meticulously designed to ensure data integrity during the saving process and data consistency during recovery.
Initially, all of these buffers are in the snapshot status.
Each snapshot process, initiated by the asynchronous thread within each training process, involves the transfer and serialization of data from the GPU into one of these snapshot buffers. 
Upon the completion of a snapshot process—and in the absence of an ongoing persist buffer—the corresponding buffer transitions to the persist status, starting the transfer of data to persistent storage.
Following the completion of the persist process, the buffer then becomes a recovery buffer, reflecting the latest checkpoint available for recovery, until another persist buffer transitions.

\subsection{Adaptive Configuration for Two-Level PEC}

Existing studies minimize $O_{lost}$ by reducing the checkpointing interval $I_{ckpt}$ \cite{mohan2021checkfreq,wang2023gemini,eisenman2022check}.
This approach may increase $O_{save}\frac{I_{total}}{I_{ckpt}}$, necessitating a considered trade-off between the two metrics.
In addition to $I_{ckpt}$, we introduce two new adjustable hyperparameters, $K_{snapshot}$ and $K_{persist}$, aimed at reducing the durations of snapshot and persist, respectively, presenting a new trade-off between efficiency and PLT.

To navigate these trade-offs across various software and hardware scenarios, we propose an adaptive configuration scheme for two-level PEC. 
Our primary strategy involves optimizing the value of $K_{snapshot}$ for snapshot-PEC to ensure it can be completely overlapped by the next F\&B, thereby minimizing $O_{save}$ while achieving a low PLT.
Even though the persist process can be fully overlapped with the subsequent training, its duration determines the lower bound for $I_{ckpt}$.
As our two-level recovery method significantly reduces the PLT caused by persist-PEC, $K_{persist}$ can be set to a relatively small value, which in turn, achieves the lowest $I_{ckpt}$.

\begin{figure*}
\centering
\begin{minipage}{1\linewidth}
    \subfigure[Total Checkpoint Size]{\includegraphics[width=0.238\linewidth]{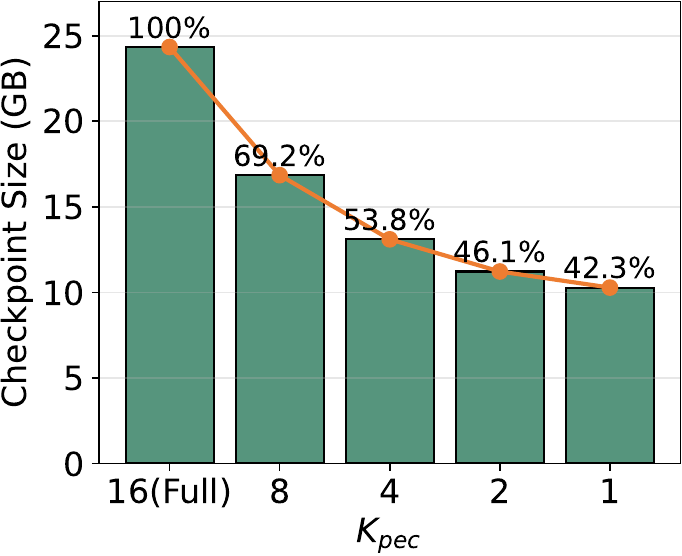}\label{fig:result_size_a}}
    \hspace{0.015in}
    \subfigure[Bottleneck Rank in Case1]{\includegraphics[width=0.248\linewidth]{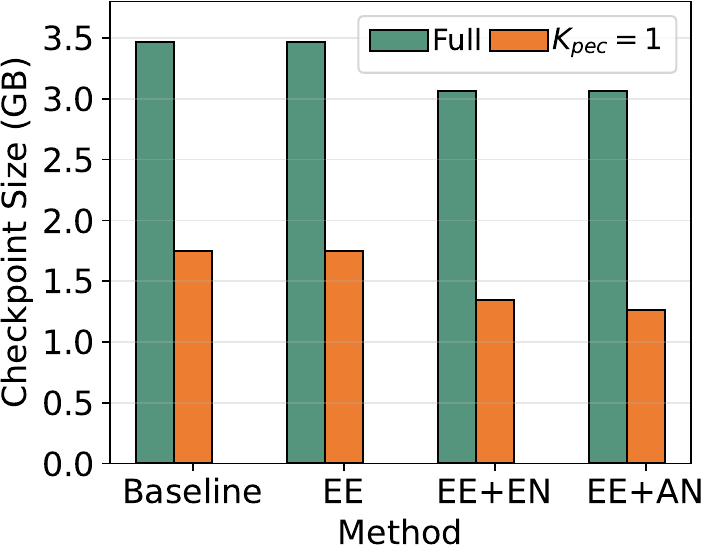}\label{fig:result_size_b}}
    \subfigure[Bottleneck Rank in Case2]{\includegraphics[width=0.248\linewidth]{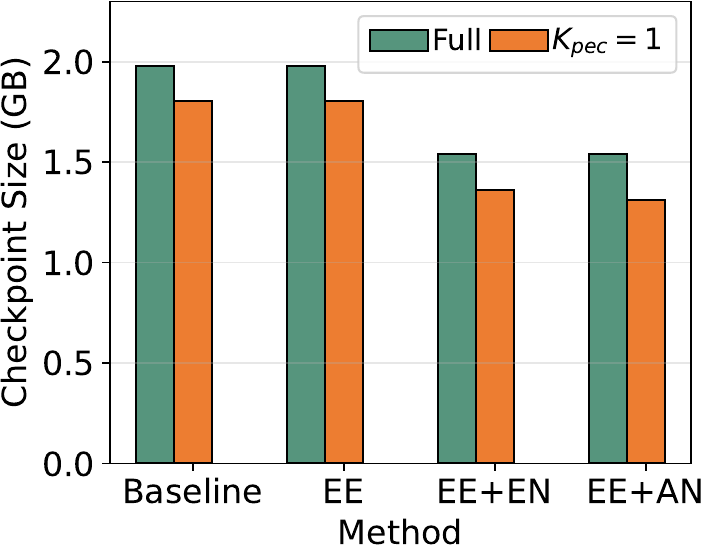}\label{fig:result_size_c}}
    \subfigure[Bottleneck Rank in Case3]{\includegraphics[width=0.248\linewidth]{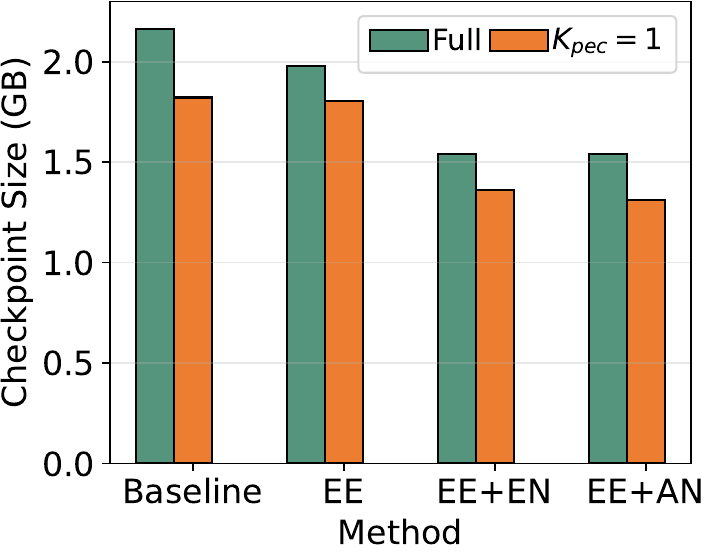}\label{fig:result_size_d}}
\end{minipage}
\vspace{-0.2in}
\caption{Experimental results of checkpoint size. (a) shows the impact of PEC on total checkpoint size. (b-d) illustrate the checkpointing workload of the bottleneck rank across various distributed configurations. ``EE'' indicates equal sharding for the expert part, while ``EN'' and ``AN'' represent equal sharding and adaptive sharding for the non-expert part, respectively.}
\label{fig:result_size}   
\end{figure*}

Additionally, as discussed in Section~\ref{sec:adaptive_sharding}, PEC may lead to workload imbalances across distributed ranks. 
In some cases, an increased $K_{pec}$ value may leverage spare capacity, reducing the PLT while maintaining the same total overhead.

\textbf{Dynamic-K for Fault Accumulation.} 
In practical scenarios, large-scale distributed training may encounter numerous faults, potentially leading to an augmented PLT.
To mitigate this issue, we propose the Dynamic-K strategy, which adjusts the $K_{pec}$ parameter in reaction to the accumulation of faults, aiming to keep PLT below a 3.75\% threshold. 
This method recalibrates the $K_{pec}$ value subsequent to each fault recovery incident, based on the aggregated PLT incurred by the system. 
If the aggregated PLT attributable to a specific $K_{pec}$ surpasses its limit, $K_{pec}$ will be doubled, and this process is reiterated until checkpointing all experts.

\begin{table}[h]
\centering
\vspace{-0.05in}
\caption{Hyperparameters for experimental MoE models.}
\label{tab:model_hyperparameters}
\vspace{-0.2in}
\begin{center}
\begin{small}
\resizebox{1\linewidth}{!}{
\begin{tabular}{lrrr}
\toprule
Parameter & GPT-125M-8E & GPT-350M-16E & SwinV2-MoE \\ 
\midrule
Num. layers & 12 & 24 & [2, 2, 18, 2]\\
Hidden size & 768 & 1024 & 96\\
Num. atten. heads & 12 & 16 & [3, 6, 12, 24]\\
Num. MoE layers & 6 & 12 & 10 \\
Num. experts/layer & 8 & 16 & 8 \\
Num. parameters & 323M & 1.7G & 173M\\
\bottomrule
\end{tabular}
}
\end{small}
\end{center}
\end{table}

\begin{table}[h]
\centering
\vspace{-0.15in}
\caption{Configurations for GPT-350M-16E model training.}
\label{tab:configurations}
\vspace{-0.2in}
\begin{center}
\begin{small}
\resizebox{1\linewidth}{!}{
\begin{tabular}{lccccccc}
\toprule
Configuration & Node & GPU & DP & TP & PP & EP & Experts/GPU \\ 
\midrule
Case1 & 1 & 8 & 8 & 1 & 1 & 8 & 2\\
Case2 & 2 & 16 & 16 & 1 & 1 & 16 & 1\\
Case3 & 2 & 16 & 16 & 1 & 1 & 8 & 2\\
\bottomrule
\end{tabular}
}
\end{small}
\end{center}
\vspace{-0.15in}
\end{table}

\section{Evaluation}
\label{sec:Eval}
\subsection{Experimental Setup}
\label{Experimental_setup}

We implement our proposed MoC-System and conduct extensive experiments upon the Megatron-DeepSpeed \cite{Megatron_DeepSpeed,smith2022using,rajbhandari2022deepspeed}, which is an acclaimed open-source framework supporting the distributed training of MoE models.
As shown in Table~\ref{tab:model_hyperparameters}, we experiment with both language and vision models.
The experimental language models (GPT-125M-8E and GPT-350M-16E) are extensions of GPT-3 like NLG model \cite{brown2020language}, provided by DeepSpeed-MoE \cite{rajbhandari2022deepspeed}.
The GPT-125M-8E model is pre-trained on the Wikitext-2 dataset \cite{wikitext} for the correlation analysis between PLT and final trained accuracy, as illustrated in Figure~\ref{fig:plt}. 
The GPT-350M-16E model is pre-trained on a 1B token subset of the SlimPajama-627B dataset \cite{cerebras2023slimpajama}.
Moreover, the distributed training of the GPT-350M-16E model is experimented with the three different configurations, as shown in Table~\ref{tab:configurations}.
Using the hybrid parallel strategy of ZeRO-2 DP \cite{rajbhandari2020zero} and EP, the training is deployed on a cluster that comprises a total of 60 nodes with 8×A800-SXM4-80GB GPUs each.
The vision model, SwinV2-MoE \cite{hwang2023tutel,swin}, is trained on the ImageNet-1K dataset \cite{deng2009imagenet}.

\begin{figure*}
\centering
\begin{minipage}{1\linewidth}
    \subfigure[Case1 Configuration]{\includegraphics[width=0.33\linewidth]{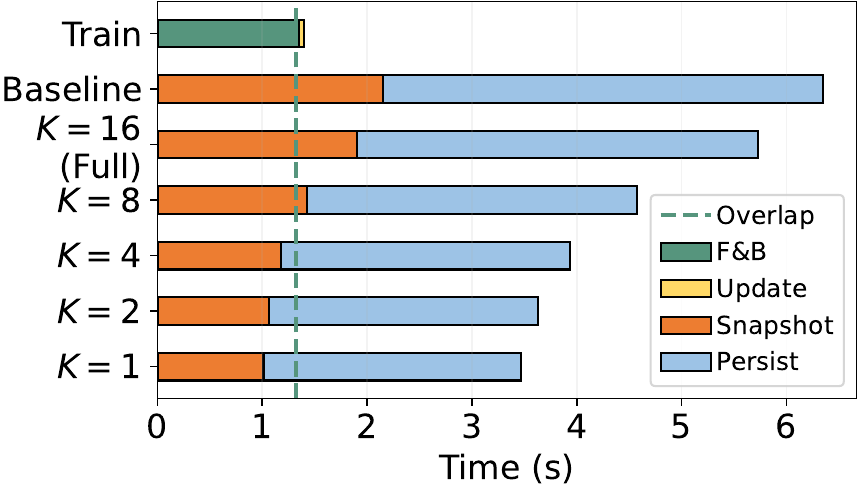}}
    \subfigure[Case2 Configuration]{\includegraphics[width=0.33\linewidth]{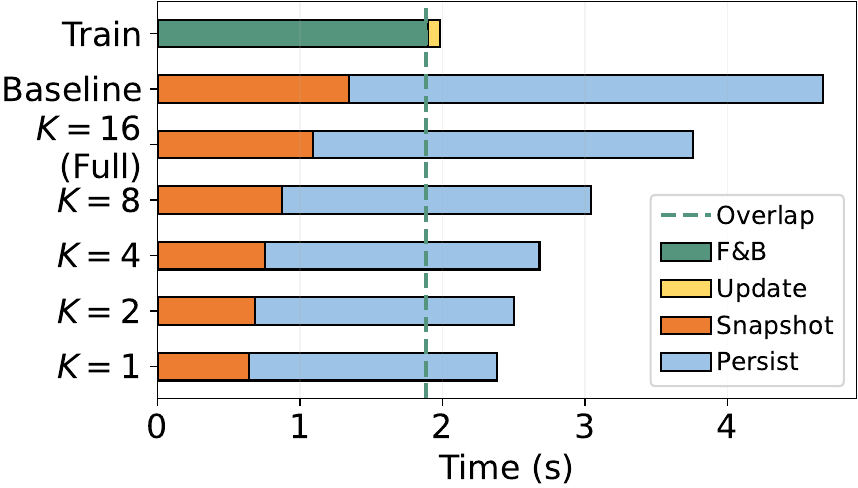}}
    \subfigure[Case3 Configuration]{\includegraphics[width=0.33\linewidth]{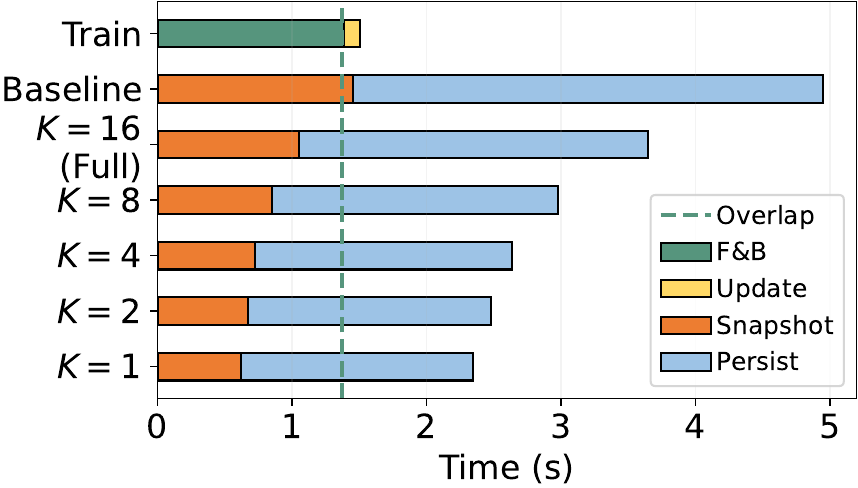}}
\end{minipage}
\vspace{-0.2in}
\caption{Duration of each process in a training iteration with checkpointing. Different values of ``$K$'' represent experiments where both $K_{snapshot}$ and $K_{persit}$ of two-level checkpointing (``Snapshot'' and ``Persist'') are set to ``$K$''. The green ``Overlap'' line marks the duration that can be overlapped by the forward and backward passes (``F\&B''). ``Update'' denotes the weight update. }
\label{fig:result_time}   
\end{figure*}

\subsection{Improvements in Checkpointing Efficiency}
\label{results_and_analysis}

\subsubsection{Checkpoint Size}
We evaluate the effectiveness of PEC in reducing checkpoint size through experiments on the GPT-350M-16E model training. 
Unless otherwise specified in subsequent experiments, PEC employs the sequential selection strategy, and the baseline refers to the method provided by the Megatron-DeepSpeed framework \cite{Megatron_DeepSpeed}. 
As illustrated in Figure~\ref{fig:result_size_a}, the total checkpoint size for each process decreases as $K_{pec}$ decreases, reaching 42.3\% of the full model checkpoint size when $K_{pec}$ is set to 1.

However, merely reducing the total checkpoint size is insufficient for optimizing efficiency in distributed training. 
As the checkpointing workload is distributed across various training ranks, the duration of the blocking checkpointing process is primarily determined by the bottleneck rank, which has the heaviest workload and longest processing time. 
Therefore, we further assess the checkpointing workload of the bottleneck rank using different sharding strategies, as illustrated in Figure~\ref{fig:result_size_b}-\ref{fig:result_size_d}.

The results indicate that our fully sharded checkpointing strategy, which applies equal sharding to both non-expert and expert parts, significantly reduces the workload of bottleneck rank in both full saving (12\% to 28\%) and PEC scenarios (22\% to 29\%). 
Notably, equal sharding of the expert part is only effective in scenarios with multiple EP groups (Case 3). 
With $K_{pec}=1$, adaptive sharding of the non-expert part can further reduce the workload by 3.7\% to 6.1\%. 

\subsubsection{Checkpointing time}
Our optimizations effectively reduce checkpoint size and balance the workload across distributed ranks, resulting in a corresponding decrease in checkpointing duration by up to about 50\%. 
In the experiments depicted in Figure~\ref{fig:result_time}, our methods all employ the fully sharded checkpointing strategy, enabling even the full savings ($K=16$) to outperform the baseline.

As discussed in Section~\ref{sec:background_async}, the snapshot process must be completely covered by the forward and backward passes in the subsequent iteration; otherwise, the weight update will be blocked. 
Figure~\ref{fig:result_time} indicates that the baseline snapshot duration exceeds the forward and backward time in Case1 and Case3. 
To address this problem, Case1 needs to employ a fully sharded checkpointing strategy, while Case3 requires saving fewer than four experts with PEC.

Moreover, the training process in Case3 is 0.5 seconds faster than in Case2, highlighting why the prevailing hybrid strategy confines EP within a node—limiting All-to-All communication to intra-node operations is more efficient than inter-node communication. 
These experiments demonstrate the broad applicability of our methods in practical scenarios.

\subsubsection{Asynchronous Checkpointing}
\begin{figure}
    \centering
    \includegraphics[width=1\linewidth]{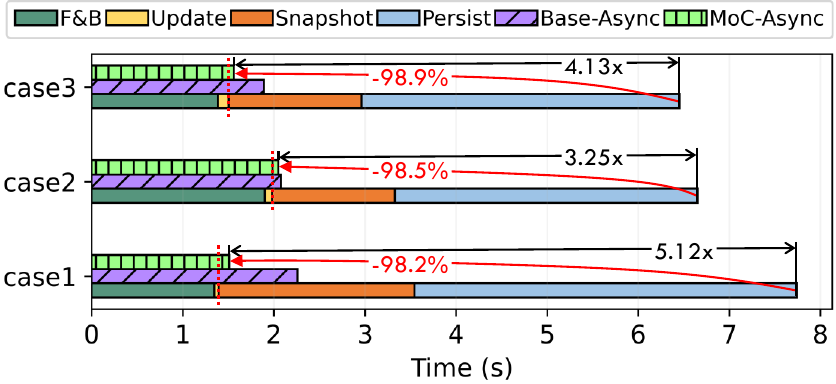}
    \vspace{-0.3in}
    \caption{Duration of a training iteration across three configurations, utilizing three checkpointing methods: (1) baseline, (2) ``Base-Async,'' which uses basic asynchronous checkpointing without our PEC and fully sharded checkpointing techniques, and (3) ``MoC-Async,'' representing the fully optimized asynchronous process within our MoC-System. \revise{``MoC-Async'' can reduce checkpointing overhead by more than 98\% compared to the baseline.}}
    \vspace{-0.1in}
    \label{fig:result_async}
\end{figure}

Given that our approaches have been validated to reduce the duration of the checkpointing process, we further assess the end-to-end optimization efficacy of our MoC-System, which implements an asynchronous checkpointing process. 
As illustrated in Figure~\ref{fig:result_async}, the fully optimized asynchronous process in our MoC-System (``MoC-Async'') can decrease the overhead of each checkpointing process ($O_{save}$) by 98.2\% to 98.9\% and accelerate each training iteration by 3.25 to 5.12 times in the three experimental cases, compared to the baseline using blocking checkpointing.
\revise{
$O_{save}$ refers to the additional time that surpasses the normal training processes, including ``F\&B'' and ``Update,'' as indicated by the duration beyond the red dotted line in Figure~\ref{fig:result_async}.
}

When our asynchronous checkpointing is applied without optimization by PEC and fully sharded techniques (``Base-Async''), it can overlap 97.9\% of the checkpointing time in Case 2, as the snapshot duration is sufficiently short to achieve complete overlap.
However, this method can only overlap 86.3\% and 92.1\% in Case 1 and Case 3, respectively, because their durations are too long to be fully overlapped. 
Employing all optimizations, ``MoC-Async'' can achieve 1.4\% to 33.2\% improvements over ``Base-Async'' in the three cases.

In addition to reducing $O_{save}$, our ``MoC-Async'' achieves half the $I_{ckpt}$ compared to the ``Base-Async'' method, as it takes only half the time to complete the snapshot and persist process. 
For instance, it reduces $I_{ckpt}$ from 2.3 to 1.2 in Case 2. 
Consequently, our ``MoC-Async'' can minimize the overall checkpoint overhead $O_{ckpt}$.

\begin{figure}
    \centering
    \includegraphics[width=1.1\linewidth]{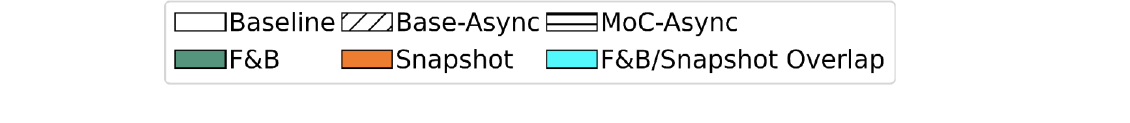}
    \vskip -0.18in
    \begin{minipage}{1\linewidth}
    \subfigure[DP+EP (A800)]{\includegraphics[width=0.495\linewidth]{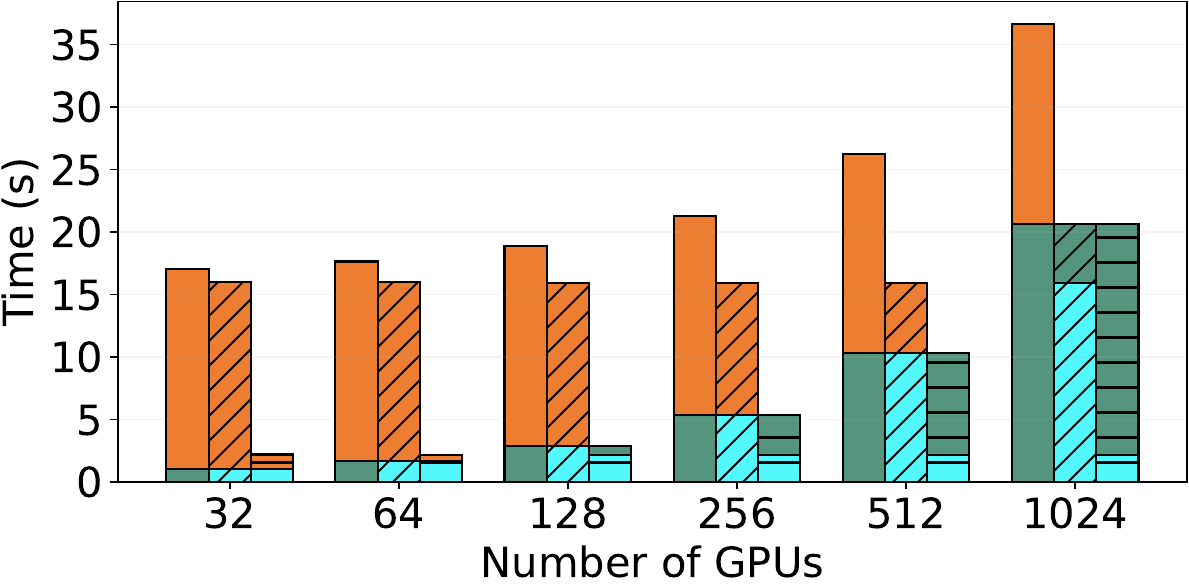}\label{fig:scaling_a}}
    \subfigure[DP+EP+TP (A800)]{\includegraphics[width=0.495\linewidth]{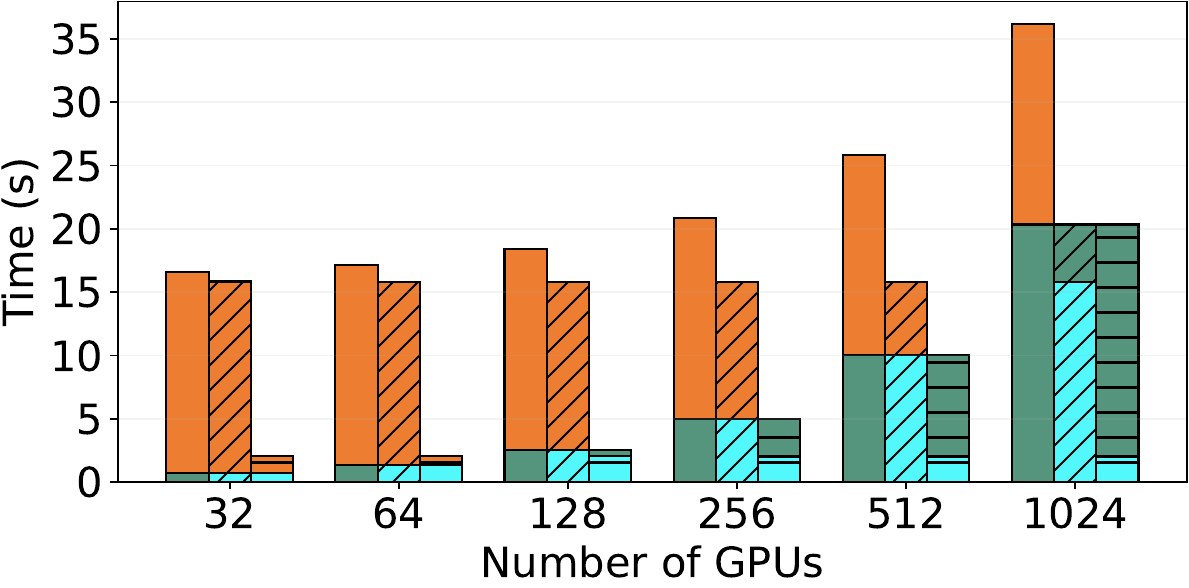}\label{fig:scaling_b}}
    \end{minipage}
    \vskip -0.15in
    \begin{minipage}{1\linewidth}
    \subfigure[DP+EP (H100)]{\includegraphics[width=0.495\linewidth]{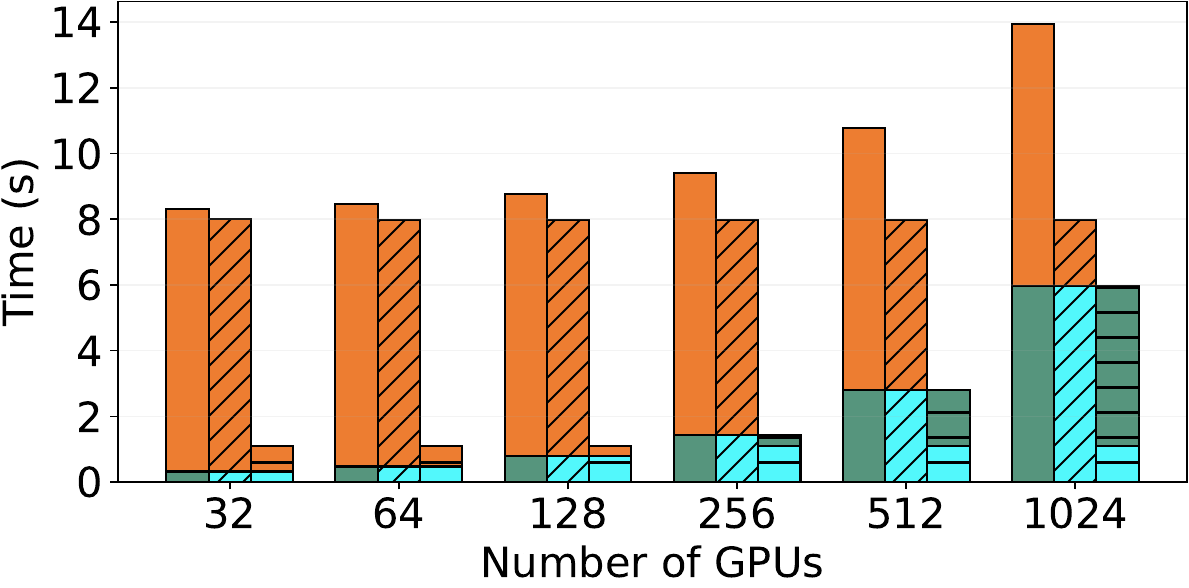}\label{fig:scaling_c}}
    \subfigure[Sequence Length]{\includegraphics[width=0.495\linewidth]{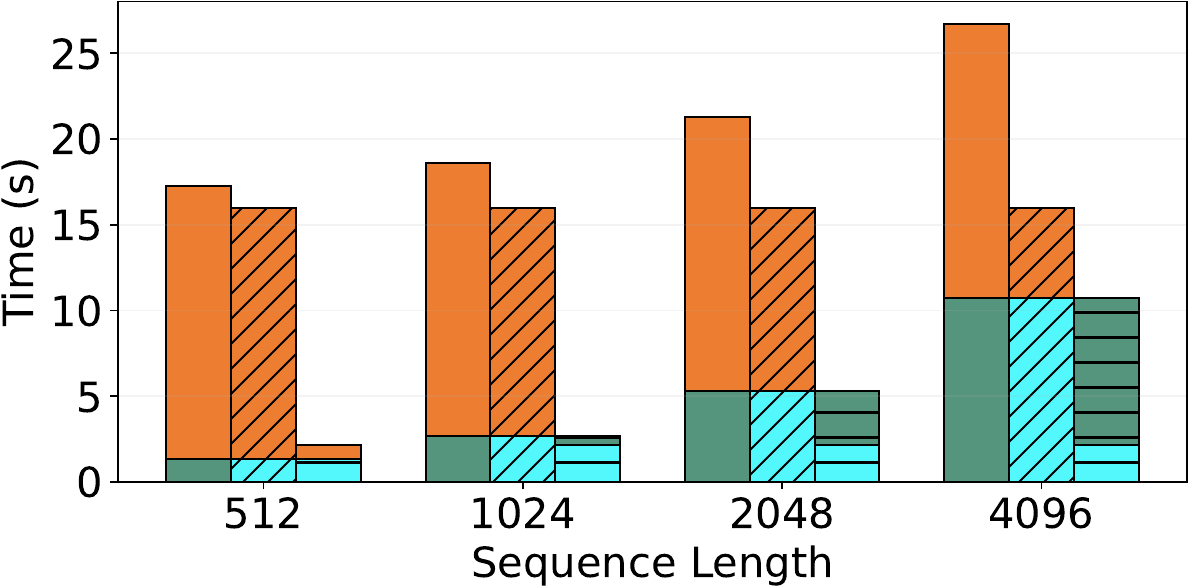}\label{fig:scaling_d}}
    \end{minipage}
    \vskip -0.15in
    \begin{minipage}{1\linewidth}
    \subfigure[Model Size]{\includegraphics[width=0.495\linewidth]{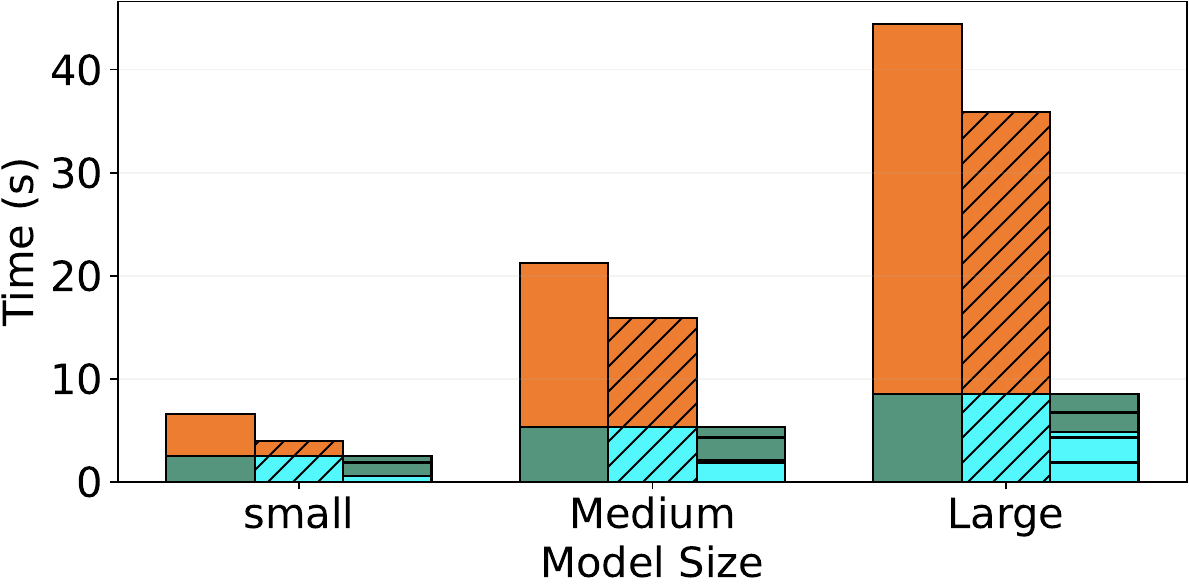}\label{fig:scaling_e}}
    \subfigure[Pesist Size]{\includegraphics[width=0.495\linewidth]{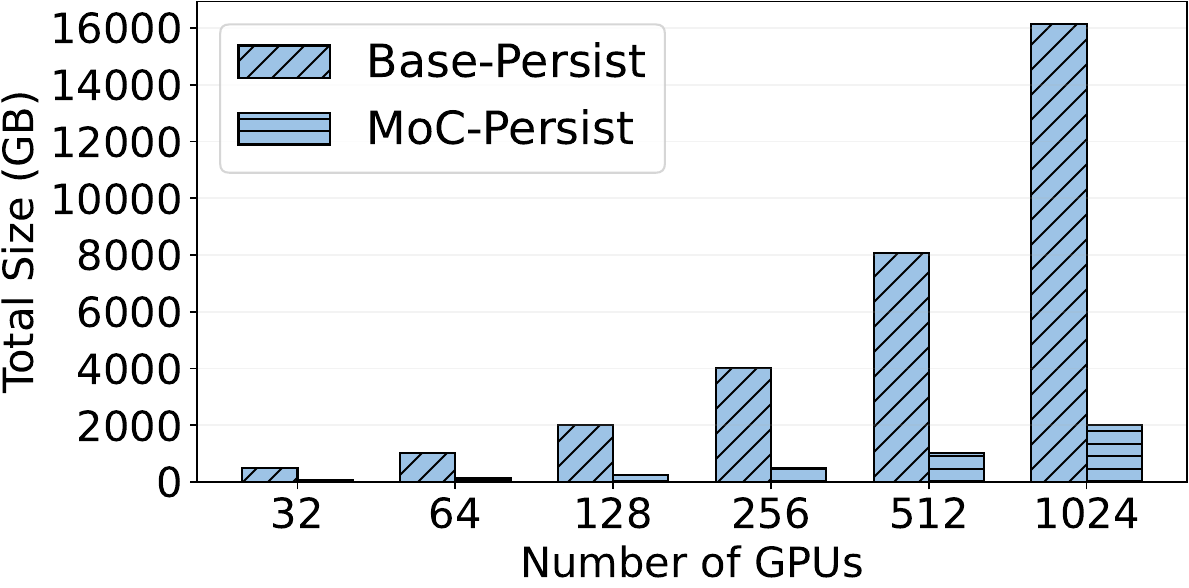}\label{fig:scaling_f}}
    \end{minipage}
    \vspace{-0.23in}
    \caption{Scaling and generalizing results across various training factors. (a) Scaling the number of A800 GPUs using DP+EP parallelism, with each GPU assigned a unique expert for each MoE layer.
(b) Scaling the number of A800 GPUs using DP+EP+TP parallelism, incorporating a 4-degree TP into the existing DP+EP configuration.
(c) Scaling the number of H100 GPUs using DP+EP parallelism.
(d) Generalizing different training sequence lengths.
(e) Generalizing different model sizes.
(f) The file size of the persist process across the DP+EP configurations with varying numbers of A800 GPUs.}
    \vspace{-0.07in}
    \label{fig:scaling}
\end{figure}

\subsubsection{Scaling and Generalizing}
\label{sec:scaling}

To demonstrate the scalability and generalizability of our proposed MoC-System, we conduct simulations to assess its efficacy across various configurations of training factors, including the number of GPUs, parallelism, hardware capability, sequence length, and model size. 
Our simulations utilize the ASTRA-SIM simulator \cite{rashidi2020astra} to model the computation and communication time costs for each iteration of distributed MoE model training across various scenarios with differing computing power and communication bandwidth. Aligned with our actual measured performance, we configure the A800 GPU simulations with 312 TFLOPS at a 20\% utilization rate and a GPU-to-CPU snapshot bandwidth of 1 GB/s. Similarly, the H100 GPU simulations are set with 989 TFLOPS at a 20\% utilization rate and a GPU-to-CPU snapshot bandwidth of 2 GB/s.

To reflect current large-scale training practices, we simulate the training of large LLaMA-like MoE models, which are commonly used in real-world applications \cite{jiang2024mixtral,liu2024deepseek}. 
In our simulations, depicted in Figure~\ref{fig:scaling}, we configure the MoE models with a hidden size of 2048, 16 attention heads, a head dimension of 128, an expert intermediate size of four times the hidden size, and 24 layers. 
Consistent with the three configurations used in Figure~\ref{fig:result_async}, Figure~\ref{fig:scaling} illustrates the duration of a training iteration of three checkpointing methods: ``Baseline,'' ``Base-Async,'' and ``MoC-Async.'' 
Furthermore, Figure~\ref{fig:scaling} breaks down the timing of the two asynchronous checkpointing methods to demonstrate the overlap duration of ``F\&B'' and ``Snapshot'', termed ``F\&B/Snapshot Overlap.''

\textbf{Number of GPUs.}
To demonstrate the scalability of the MoC-System for large-scale training, we scale the model training across varying numbers of GPUs. 
Specifically, we employ Data Parallelism (ZeRO-2) and Expert Parallelism by assigning each expert of an MoE layer to a distinct GPU, scaling both the system and the model size. 
Figure~\ref{fig:scaling_a} shows that the ``F\&B,'' which can be used to overlap the snapshot overhead, significantly increases as the number of GPUs increases. 
\revise{
Compared to the baseline, the two asynchronous checkpointing methods effectively facilitate overlap, thereby reducing time costs.
When the number of GPUs is less than 1024, the snapshot duration of ``Base-Async'' is too long to be fully overlapped, resulting in lower efficiency compared to ``MoC-Async.'' 
In contrast, ``MoC-Async,'' configured to save only 1/8 of the experts per checkpoint, reduces the required snapshot time, making ``F\&B'' the bottleneck of the total time when exceeding 64 GPUs.
Additionally, ``MoC-Async'' can achieve substantial optimization even when ``F\&B'' is considerably less than ``Snapshot,'' a benefit that ``Base-Async'' is unable to provide.
}

In this setup, an equal number of distinct expert parameters is allocated to each GPU, resulting in the data volume for each GPU-to-CPU snapshot remaining similar.
However, as the number of GPUs increases, the data volume for CPU-to-storage persist on the cluster filesystem grows significantly, as illustrated in Figure~\ref{fig:scaling_f}. 
Our ``MoC-Persist'' method significantly reduces the persist file size and the time required for the persist process, enabling shorter checkpoint intervals and consequently minimizing the lost time due to faults.

\textbf{Parallelism.}
To generalize the efficacy of the MoC-System across various parallelism configurations, we further investigate training using the DP+EP+TP parallelism to train models with the same number of experts as in the DP+EP scenario, as depicted in Figure~\ref{fig:scaling_b}. 
Although different parallelism strategies impact the iteration time of ``F\&B,'' the behavior observed during checkpointing is similar to that seen with the DP+EP configuration (Figure~\ref{fig:scaling_a}).
\revise{
Consistently, ``MoC-Async'' maintains optimal efficiency across all tested GPU configurations, particularly when the number of GPUs is fewer than 1024, where the snapshot duration of ``Base-Async'' cannot be fully overlapped.
}

\textbf{Hardware Capability.}
To generalize the efficacy of the MoC-System across different hardware platforms, we conduct training simulations with the A800 GPU (Figure~\ref{fig:scaling_a}) and the H100 GPU (Figure~\ref{fig:scaling_c}) configurations. 
Due to variations in capabilities such as GPU computing power, GPU memory bandwidth, NVLink bandwidth, and GPU-to-CPU PCIe bandwidth, the durations of both ``F\&B'' and ``Snapshot'' differ in the H100 scenario, resulting in varying overlap performance. 
\revise{
Specifically, ``MoC-Async'' demonstrates significantly greater efficiency than other methods across all tested H100 configurations, as the snapshot of ``Base-Async'' cannot be fully overlapped even with 1024 GPUs.
It is anticipated that the computation and communication capabilities associated with ``F\&B'' will advance more rapidly than the GPU-to-CPU data transmission capabilities in future hardware platforms. Therefore, reducing the data volume of checkpointing through ``MoC-Async'' will remain valuable for larger-scale distributed training in the future.
}

\textbf{Sequence Length.}
Sequence length is a critical factor influencing the ``F\&B'' duration. 
To investigate its impact, we conduct simulations with varying sequence lengths using the DP+EP configuration on 256 A800 GPUs. 
While longer sequences significantly increase the ``F\&B'' time, variations in sequence length do not impact the checkpointing process, as shown in Figure~\ref{fig:scaling_d}.
This is because the checkpointed data pertains to the constant model parameters rather than the dynamic activations. 
\revise{
Consequently, ``MoC-Async'' can achieve higher efficiency across all evaluated sequence lengths.
}

\textbf{Model Size.}
As larger model sizes lead to increased iteration times and larger data volumes for checkpointing, we conduct simulations using models of three different sizes: a hidden size of 1024 (``Small''), 2048 (``Medium''), and 3072 (``Large''), as illustrated in Figure~\ref{fig:scaling_e}. These simulations are carried out using the DP+EP parallelism with 256 A800 GPUs. 
The results show that ``MoC-Async'' not only improves efficiency across various model sizes but provides more pronounced efficacy in scenarios involving larger-scale models.
\revise{
This is because model size impacts both the ``F\&B'' and ``Snapshot.''
Due to the disparity in capabilities between computation and GPU-to-CPU data transmission, the duration of snapshots increases more significantly with the growth of the model size.
}

\begin{table*}[h]
\caption{Accuracy results (\%) of the models on downstream tasks, pre-trained as shown in Figure~\ref{fig:accuracy_result1}. The downstream tasks includes: HellaSwag \cite{zellers2019hellaswag}, PIQA \cite{bisk2020piqa}, WinoGrande \cite{sakaguchi2021winogrande}, BoolQ \cite{clark2019boolq}, ARC-Easy \cite{clark2018think}, OBQA \cite{OpenBookQA2018}, RACE \cite{lai-etal-2017-race}, MathQA \cite{amini2019mathqa}. ``Ckpt'' indicates the relative total checkpoint size compared to the baseline, which saves the full model states. ``Deviation'' shows the deviation of the minimum and maximum accuracy of our methods from the baseline.}
\label{tab:downstream}
\vspace{-0.15in}
\begin{center}
\begin{small}
\resizebox{1\linewidth}{!}{
\begin{tabular}{l|c|cccccccc|c}
\toprule
\textbf{Method} & \textbf{Ckpt} & \textbf{HellaSwag} & \textbf{PIQA} & \textbf{WinoGrande} & \textbf{BoolQ} & \textbf{ARC-E} & \textbf{OBQA} & \textbf{RACE} & \textbf{MathQA} & \textbf{Avg. ($\uparrow$)}\\
\midrule
Baseline & 1 & 26.85&	58.22&	49.09&	54.77& 36.83& 13.00& 24.21&	20.54&	35.44 \\
W & 0.88 & 26.92&	58.16&	49.72&	57.52& \textbf{37.84}& 12.80&	24.69&	\textbf{20.84}&	36.06 \\
O & 0.54 & 26.93&	58.00&	48.54&	61.28& 37.21& \textbf{13.40}& \textbf{25.26}& 19.97& 36.32 \\
WO & 0.42 & 26.91&	58.38&	49.33&	61.31& 37.33&	13.20&	24.50&	20.20&	36.40 \\
WO-2L & 0.42 & \textbf{26.96}&	\textbf{58.49}&	\textbf{50.12}&	\textbf{61.74}&	37.12&	13.20&	24.40&	20.13&	\textbf{36.52} \\
\midrule
Deviation & -	& (0.06, 0.11)&	(-0.22, 0.27) &	(-0.55, 1.03) &	(2.75, 6.97) & (0.29, 1.01)& (-0.20, 0.40)& (0.19, 1.05) &	(-0.57, 0.30)&	(0.62, 1.08) \\
\bottomrule
\end{tabular}
}
\end{small}
\end{center}
\end{table*}

\subsubsection{Modeling and Analysis} 
Based on the scaling and generalizing simulations detailed in Section~\ref{sec:scaling}, we conclude that the identified factors influence the efficiency of the checkpointing system in two ways: (1) by affecting the duration of each iteration ($T_{F\&B}$), and (2) by affecting the time required for the snapshot ($T_{Snapshot}$). 
Specifically, sequence length affects only $T_{F\&B}$, whereas the number of GPUs, parallelism, hardware capability, and model size influence both $T_{F\&B}$ and $T_{Snapshot}$.
Together, these factors determine the checkpoint saving overhead ($O_{save}$), ideally expressed as:
\begin{equation}
\begin{aligned}
\label{eq:analysis_0}
O_{save} = \begin{cases} 
T_{Snapshot} - T_{F\&B}, & \text{if } T_{Snapshot} > T_{F\&B} \\
0. & \text{if } T_{Snapshot} \leq T_{F\&B} 
\end{cases}
\end{aligned}
\end{equation}

Based on Equation~\ref{eq:overhead_2} in Section~\ref{sec:overhead} and assuming a constant failure rate denotes as $\lambda$, the number of faults can be
\begin{equation}
\begin{aligned}
\label{eq:analysis_1}
  N_{fault} \approx \lambda I_{total}. 
\end{aligned}
\end{equation}
The total overhead associated with the existing full checkpointing method, which saves all model states and is denoted as $O_{ckpt}^{Full}$, as well as our proposed MoC method, denoted as $O_{ckpt}^{MoC}$, can be expressed as follows:
\begin{equation}
\begin{aligned}
\label{eq:analysis_2}
O_{ckpt}^{Full} \approx O_{save}^{Full}\frac{I_{total}}{I_{ckpt}^{Full}} + \lambda I_{total} \left(O_{restart} + \frac{I_{ckpt}^{Full}}{2}\right),
\end{aligned}
\end{equation}
\begin{equation}
\begin{aligned}
\label{eq:analysis_3}
O_{ckpt}^{MoC} \approx O_{save}^{MoC}\frac{I_{total}}{I_{ckpt}^{MoC}} + \lambda I_{total} \left(O_{restart} + \frac{I_{ckpt}^{MoC}}{2}\right).
\end{aligned}
\end{equation}
$O_{save}^{Full}$ and $O_{save}^{MoC}$ represent the overhead associated with saving model states for two respective methods, while $I_{ckpt}^{Full}$ and $I_{ckpt}^{MoC}$ denote the checkpointing intervals configured for each method.
To ensure that the overhead of the MoC method is less than that of the full checkpointing method, the following conditions must be met:
\begin{equation}
\begin{aligned}
\label{eq:analysis_4}
O_{ckpt}^{MoC} < O_{ckpt}^{Full},
\end{aligned}
\end{equation}
\begin{equation}
\begin{aligned}
\label{eq:analysis_5}
\frac{O_{save}^{MoC}}{I_{ckpt}^{MoC}} + \lambda \left(O_{restart} + \frac{I_{ckpt}^{MoC}}{2}\right) < \frac{O_{save}^{Full}}{I_{ckpt}^{Full}} + \lambda \left(O_{restart} + \frac{I_{ckpt}^{Full}}{2}\right),
\end{aligned}
\end{equation}
\begin{equation}
\begin{aligned}
\label{eq:analysis_6}
\frac{O_{save}^{MoC}}{I_{ckpt}^{MoC}} + \lambda \frac{I_{ckpt}^{MoC}}{2} < \frac{O_{save}^{Full}}{I_{ckpt}^{Full}} + \lambda \frac{I_{ckpt}^{Full}}{2}. 
\end{aligned}
\end{equation}

Based on the condition outlined in Equation~\ref{eq:analysis_6}, we can identify two strategies to leverage the advantages of our proposed MoC design: (1) By maintaining the same checkpoint interval as the full checkpointing method, MoC can reduce overhead because it has a smaller saving overhead for each checkpointing. (2) By decreasing the checkpoint interval (more frequent checkpointing) to equalize the ratio of $O_{save}$ to $I_{ckpt}$ between MoC and the full method, MoC can still reduce the total overhead. This reduction is achieved by minimizing the lost time caused by faults, as the lost time is directly proportional to the checkpoint interval.

\begin{figure}
\centering
\vspace{-0.05in}
\begin{minipage}{1\linewidth}
    \subfigure[GPT-350M-16E]{\includegraphics[width=0.491\linewidth]{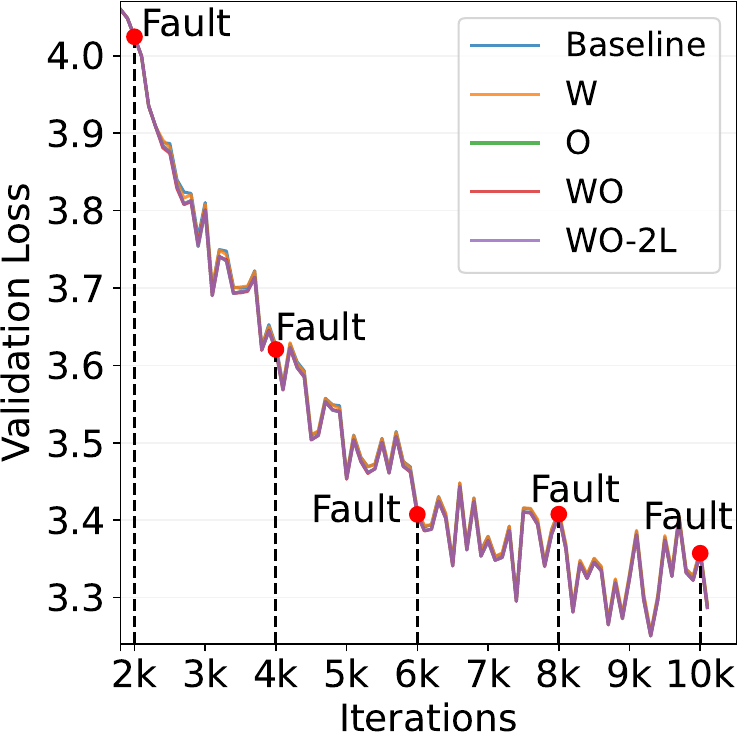}\label{fig:accuracy_result1_a}}
    \subfigure[SwinV2-MoE]{\includegraphics[width=0.502\linewidth]{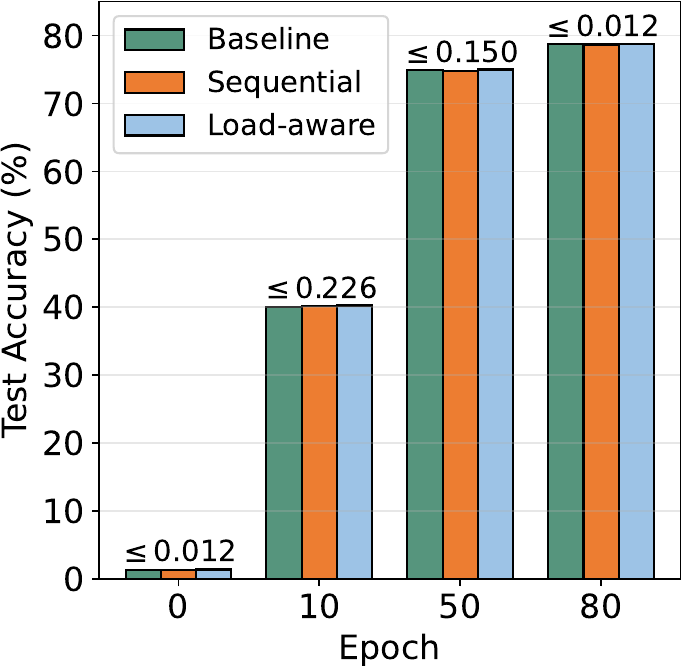}\label{fig:accuracy_result1_b}}
\end{minipage}
\vspace{-0.2in}
\caption{Loss curve of the GPT-350M-16E model pre-training (a) and test accuracy of the SwinV2-MoE model pre-training (b). In (a), faults occur every 2k iterations, indicated by red points. ``W'' and ``O'' denote the use of PEC ($K_{snapshot}=4$ and $K_{persist}=1$) on weights and optimizer states, respectively. ``-2L'' refers to the use of two-level recovery, while methods without ``-2L'' recover solely from the persistent checkpoint stored in storage. In (b), faults are introduced at epochs(0, 10, 50, 80), with test accuracy reported at the conclusion of each epoch. }
\vspace{-0.05in}
\label{fig:accuracy_result1}   
\end{figure}

\subsection{Impact on Model Accuracy}
\label{sec:eval_accuracy}

In Section~\ref{sec:PEC_analysis_accuracy} and Figure~\ref{fig:plt}, we initially demonstrate the efficacy of our proposed PEC in reducing checkpoint size without compromising model accuracy. 
We then conduct an in-depth evaluation of its impact on model accuracy.

As shown in Figure~\ref{fig:accuracy_result1_a}, applying PEC to save model weights (``W''), optimizer states (``O''), or both (``WO'' and ``WO-2L'') results in a validation loss curve comparable to the baseline, which saves the full states during the pre-training of the GPT-350M-16E model.

Given the similar training curves across different checkpointing methods, we further evaluate downstream tasks for each pre-trained model. 
Compared to the baseline method, which retains all states, our lossy methods (``W'', ``O'', ``WO'' and ``WO-2L'') achieve higher average accuracy, ranging from 0.62\% to 1.08\%, as shown in Table~\ref{tab:downstream}. 
Notably, our methods show the most significant accuracy improvement on the BoolQ task, ranging from 2.75\% to 6.97\%.
We hypothesize that this level of improvement may result from state loss caused by our PEC, acting as a variant of dropout \cite{srivastava2014dropout}, which helps prevent overfitting in certain domains.

\subsubsection{Two-level PEC Saving and Recovery.} 
We evaluate the effectiveness of our two-level PEC saving and recovery scheme in minimizing PLT and maintaining model accuracy.
Given the faster speed of the snapshot process compared to the persist process, we configure $K_{persist}=1$ and experiment with varying $K_{snapshot}$ values, as depicted in Figure~\ref{fig:twolevel_dynamic_a}. 
Compared with the baseline ($K_{snapshot}=1$, $K_{persist}=1$) setup, increasing $K_{snapshot}$ markedly reduces PLT, owing to the retrieval of partial experts from the in-memory snapshots on the non-fault node. 
Moreover, the two-level recovery with the ($K_{snapshot}=4$, $K_{persist}=1$) setup (``WO-2L'' in Table\ref{tab:downstream}) achieves the highest average accuracy on downstream tasks, exceeding the baseline by 1.08\%.


\subsubsection{Sequential versus Load-aware Selection} 
We conduct experiments on the SwinV2-MoE model pre-training to evaluate the impact of different partial expert selection methods on model accuracy. 
As shown in Figure~\ref{fig:accuracy_result1_b}, the three methods---baseline, PEC with sequential selection, and PEC with load-aware selection---exhibit minimal differences, with less than a 0.0012\% variance in test accuracy after 80 training epochs. 
Considering that load-aware selection incurs additional control and synchronization costs while maintaining comparable accuracy, sequential selection appears to be the more practical choice for real-world applications. 
Additionally, these experiments confirm that our PEC method is applicable to both language and vision models.

\subsubsection{Dynamic-K} 
\begin{figure}
\centering
\vspace{-0.03in}
\begin{minipage}{1\linewidth}
    \subfigure[Two-Level Recovery]{\includegraphics[width=0.5\linewidth]{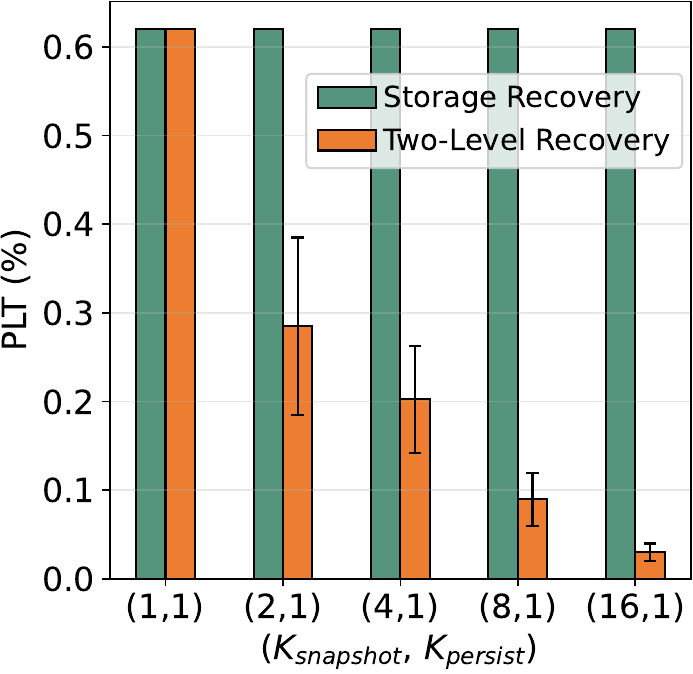}\label{fig:twolevel_dynamic_a}}
    \subfigure[Dynamic-k]{\includegraphics[width=0.480\linewidth]{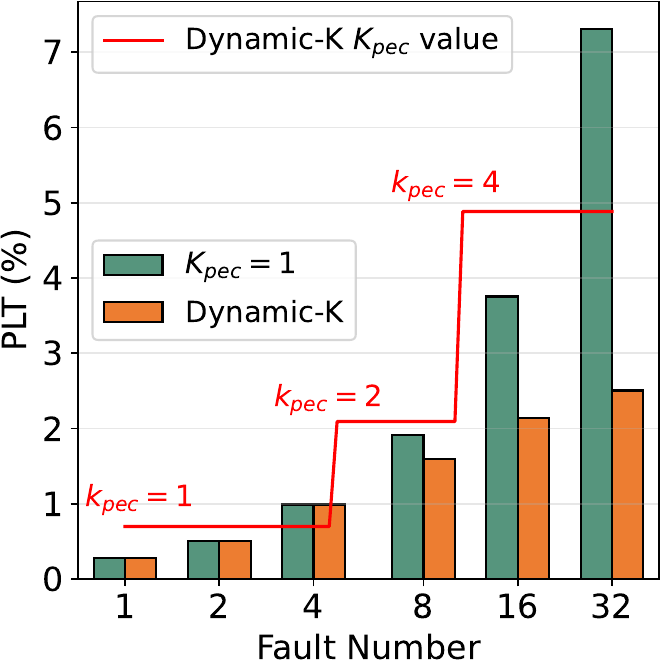}\label{fig:twolevel_dynamic_b}}
\end{minipage}
\vspace{-0.15in}
\caption{(a) shows the correlation between PLT and various combinations of $K_{snapshot}$ and $K_{persist}$, using two-level recovery. The error bar represents the fluctuation in measured values. (b) demonstrates the efficacy of our Dynamic-K strategy in reducing PLT, with the red line tracking the dynamic adjustments of $K_{pec}$. These experiments are conducted during the pre-training of the GPT-350M-16E model in Case2.}
\label{fig:twolevel_dynamic}   
\end{figure}

We evaluate the efficacy of our proposed dynamic-K strategy in ensuring that the PLT does not exceed the pre-set threshold of 3.75\% as the number of faults increases. 
As shown in Figure~\ref{fig:twolevel_dynamic_b}, the value of $K_{pec}$ dynamically adjusts from 1 to 4, in response to escalating fault occurrences.
With this strategy, the cumulative PLT remains at a low level, whereas a constant setting of $K_{pec}=1$ results in a linear increase.

\begin{table}[t]
\caption{Accuracy results from fine-tuning the OLMoE \cite{muennighoff2024olmoe} model using various methods. ``Base'' refers to the pre-trained model without fine-tuning, ``FT-w.o.E'' indicates the fine-tuned model without fine-tuning all expert parameters, ``FT-Full'' represents the fine-tuned model with full state saving at each checkpointing, and ``FT-PEC'' denotes the fine-tuned model utilizing PEC that saves 1/8 of the experts at each checkpoint. The tasks includes: HellaSwag \cite{zellers2019hellaswag}, PIQA \cite{bisk2020piqa}, WG \cite{sakaguchi2021winogrande}, BoolQ \cite{clark2019boolq}, ARC-C \cite{clark2018think}, OBQA \cite{OpenBookQA2018}, RTE \cite{wang2019superglue}.}
\label{tab:finetune}
\vspace{-0.15in}
\begin{center}
\begin{small}
\resizebox{1\linewidth}{!}{
\begin{tabular}{l|ccccccc|c}
\toprule
\textbf{Method} &  \textbf{HS} & \textbf{PIQA} & \textbf{WG} & \textbf{BQ} & \textbf{ARC} & \textbf{OBQA} & \textbf{RTE}& \textbf{Avg.}\\
\midrule
Base & 57.99& 80.52& 68.59& 74.46&	47.27&	44.80&	54.51&	61.16 \\
FT-w.o.E & 58.58&	81.88&	68.51&	76.82&	48.72&	45.20&	63.54&	63.32 \\
FT-Full & 58.34&	81.34&	70.40&	79.11&	48.38&	45.00&	66.06&	64.09 \\
FT-PEC & 58.78&	81.45&	70.24&	79.17&	48.23&	45.00&	65.58&	64.06 \\
\bottomrule
\end{tabular}
}
\end{small}
\end{center}
\vspace{-0.1in}
\end{table}

\subsubsection{Fault Tolerance during Fine-Tuning}
In addition to the model's pre-training phase, fine-tuning is another crucial stage that requires extended training periods and fault tolerance. To evaluate the impact of our proposed PEC during the fine-tuning phase, we conduct experiments using the Alpaca dataset \cite{alpaca} to fine-tune the open-source, pre-trained OLMoE model \cite{muennighoff2024olmoe}. 
We set a fault interruption occurring halfway through the process. 
As shown in Table~\ref{tab:finetune}, PEC maintains accuracy comparable to the full-saving method.
Additionally, we conduct experiments on fine-tuning with freezing all the expert parameters. 
This approach still achieves an increase in average accuracy, from 61.16\% to 63.32\%, with only a slight degradation of 0.77\% compared to full-parameter fine-tuning. 
These results further substantiate that the expert parameters are less sensitive to a limited number of updates.

\section{Conclusion \& Future Work}
The advent of MoE models poses efficiency challenges for conventional fault-tolerant checkpointing methods due to the substantial escalation in model parameters.
Breaking new ground in efficient fault tolerance for MoE model training, we propose the Mixture-of-Checkpoint System (MoC-System).
This system integrates an innovative algorithm-system co-design---Partial Experts Checkpoint (PEC) mechanism---along with multiple optimization strategies, including fully sharded checkpointing and two-level checkpointing management.
Empirical evaluations substantiate that our MoC-System significantly reduces checkpointing overhead without compromising model accuracy. 

While existing LLM checkpointing ensures algorithm invariance, the MoC-System illustrates the feasibility of a more flexible algorithm-system co-designed approach to fault tolerance. 
As system efficiency becomes increasingly important in LLM development, more algorithms are being co-designed to enhance efficiency during training and inference. 
We believe fault tolerance can also be more closely integrated with LLM algorithms.
In future work, we intend to explore features of LLMs, such as sparsity, to develop more efficient co-design strategies for LLM training and fault tolerance.




\begin{acks}
We thank the anonymous reviewers and our shepherd, Yiran Chen, for their valuable comments and suggestions. This work was supported in part by the National Key R\&D Program of China (No. 2024YFB4505800), the National Natural Science Foundation of China (No. 62402411), the Guangdong Basic and Applied Basic Research Foundation (No. 2023A1515110353), the Guangdong Provincial Talent Program (No. 2023QN10X252), and the Guangzhou-HKUST(GZ) Joint Funding Program (No. 2024A03J0624). This research was conducted on the High-Performance Computing Platform of HKUST(GZ).
\end{acks}

\bibliographystyle{ACM-Reference-Format}
\balance
\bibliography{references}


\end{document}